\pdfoutput=1
\documentclass[11pt]{article}

\usepackage[a4paper,margin=2.4cm]{geometry}
\usepackage{amsmath,amssymb,amsthm}
\usepackage{bm}
\usepackage{graphicx}
\usepackage[numbers,sort&compress]{natbib}
\usepackage{xcolor}
\usepackage{booktabs}
\usepackage{array}
\usepackage[font=small,labelfont=bf,labelsep=period]{caption}
\usepackage[colorlinks=true,linkcolor=red!70!black,citecolor=blue,urlcolor=blue]{hyperref}
\usepackage{doi}
\newtheorem{theorem}{Theorem}
\newtheorem{proposition}{Proposition}
\newtheorem{lemma}{Lemma}
\newtheorem{corollary}{Corollary}
\theoremstyle{definition}
\newtheorem{definition}{Definition}

\newcommand{\ip}[2]{\langle #1,#2\rangle}
\newcommand{\tr}{\operatorname{tr}}
\newcommand{\supp}{\operatorname{supp}}
\newcommand{\spn}{\operatorname{span}}
\newcommand{\EQ}{\mathrm{EQ}}
\newcommand{\Ind}{\mathbf{1}}
\newcommand{\Id}{{\bf I}}
\newcommand{\C}{\mathrm{C}}
\newcommand{\Q}{\mathrm{Q}}
\newcommand{\AQ}{\mathrm{AQ}}

\hypersetup{pdftitle={An Unbounded Separation of Almost-Quantum and Commuting-Operator Correlations in Zero-Error Communication},pdfauthor={R. Gazi, B. Chatterjee, J. Karmakar, R. K. Patra, M. Banik, A. Das Bhowmik}}

\begin{document}

\begin{center}
{\Large\bfseries An Unbounded Separation of Almost-Quantum and Commuting-Operator Correlations in Zero-Error Communication}

\vspace{8pt}
{Rafiuddin Gazi$^{1}$, Biswadeep Chatterjee$^{1}$, Jayashree Karmakar$^{1}$, Ram Krishna Patra$^{2}$,\\ Manik Banik$^{1}$ and Anandamay Das Bhowmik$^{1}$\par}
{\small
$^1$S. N. Bose National Centre for Basic Sciences, Block JD, Sector III, Salt Lake, Kolkata 700 106, India\\
$^2$HUN-REN Institute for Nuclear Research, PO Box 51, H-4001 Debrecen, Hungary\\
}
\end{center}

\begin{abstract}
\noindent Almost-quantum correlations form the $\mathtt{1+AB}$ level of the Navascu\'es--Pironio--Ac\'in hierarchy. They strictly contain the quantum set, are characterized by a single semidefinite program, arise in Sorkin's quantum measure theory, and satisfy several principles proposed to single out quantum theory. We show that they are nevertheless unboundedly stronger than quantum correlations in a zero-error communication task. Alice and Bob receive labels $s,t\in[N]$, $N\ge3$; after one bit from Alice, Bob must decide whether $s=t$, never erring but allowed to abstain. For every model of quantum correlations, from finite-dimensional tensor products to commuting operators, the optimal worst-case probability of a conclusive answer is exactly $1/N$, the value attained with shared randomness, so the bound does not depend on the resolution of Tsirelson's problem. An explicit family of almost-quantum correlations attains $1/19$ for every $N$. For certificates of equality from binary-outcome boxes, we solve the underlying semidefinite programs in closed form. The almost-quantum optimum equals $N^2/(17N^2-80N+64)$ for $N\ge8$ and tends to $1/17$. The first level of the hierarchy, which coincides with the set singled out by macroscopic locality, attains $N/[8(N-2)]$, which tends to $1/8$; its optimal boxes lie outside the almost-quantum set for every $N\ge5$. For longer messages, entanglement saves at most one bit over shared randomness, whereas almost-quantum correlations reduce the one-way zero-error cost of deciding equality of $n$-bit strings from $n$ bits to at most $13$. The quantum bound also yields a linear witness of post-quantumness on a face of the nonsignalling polytope.
\end{abstract}



\section{Introduction}\label{sec:intro}

Quantum theory allows distant observers to share correlations that no classical common cause can explain, and these correlations are a resource for information processing. Shared entanglement can reduce the communication needed to compute distributed functions~\cite{CleveBuhrman1997,BuhrmanRMP2010}, in some cases exponentially~\cite{BuhrmanCleveWigderson1998,BRSW2012,Chakraborty2026}. The origin of this advantage is Bell nonlocality~\cite{Bell1964,Bell1966,Mermin1993,Brunner2014}: correlations that admit a local hidden-variable model can be simulated with shared randomness, whereas violations of Bell inequalities can be converted into communication protocols that outperform every classical one~\cite{Brukner2004,BuhrmanRMP2010}. Quantum theory, however, does not exhaust the correlations compatible with relativistic causality, and stronger correlations would be more powerful still. Popescu--Rohrlich (PR) boxes~\cite{PR1994,Popescu2014}, for example, would make communication complexity trivial~\cite{vanDam2013,Brassard2006,BrunnerSkrzypczyk2009}. Observations of this kind have motivated a long-standing programme to identify physical or information-theoretic principles that rule out supra-quantum correlations and, ideally, single out the quantum set. Prominent candidates include nontrivial communication complexity~\cite{Brassard2006}, no advantage for nonlocal computation~\cite{Linden2007}, information causality~\cite{Pawlowski2009,Cavalcanti2010,Patra2023}, macroscopic locality~\cite{NavascuesWunderlich2010} and local orthogonality~\cite{Fritz2013}.

Almost-quantum (AQ) correlations~\cite{Navascues2015} are the most demanding test case for this programme. They form the ``$\mathtt{1+AB}$'' $\mathtt{Level}$ of the Navascu\'es--Pironio--Ac\'in (NPA) hierarchy~\cite{NPA2007,NPA2008}, a set that strictly contains the quantum set yet is characterized by a single semidefinite program. AQ correlations have been proved to satisfy nontrivial communication complexity, no advantage for nonlocal computation, macroscopic locality and local orthogonality~\cite{Navascues2015}, and numerical evidence suggested that they also satisfy information causality~\cite{Navascues2015}. They also arise outside the Hilbert-space formalism. In Sorkin's quantum measure theory~\cite{Sorkin1994}, a histories-based generalization of quantum mechanics, strongly positive decoherence functionals yield correlations lying between the AQ set and NPA $\mathtt{Level\,1}$, and exactly the AQ set when later measurements may depend on earlier outcomes~\cite{DowkerHensonWallden2014}. The AQ set also follows from a principle of macroscopic noncontextuality~\cite{HensonSainz2015}. Bipartite principles may in any case be insufficient to recover the quantum set~\cite{Gallego2011}, and the known arguments that exclude AQ correlations, based on Specker's principle in contextuality scenarios~\cite{Gonda2018} and on the no-restriction hypothesis for generalized probabilistic theories~\cite{Sainz2018}, rely on structure beyond correlations in Bell scenarios. Within Bell scenarios, none of the principles listed above other than information causality is known to exclude AQ correlations; that information causality does so, with the AQ boxes constructed here, is shown in a companion work~\cite{DasBhowmikBanik2026} (see the Note added at the end of Section~\ref{sec:discussion}). Is the difference between AQ and quantum correlations then also negligible in operational terms?

Here we show that it is not. We consider a simple communication task. Alice and Bob each receive one of $N$ labels, Alice sends a single bit to Bob, and Bob must decide whether their labels coincide; he may abstain, but he must never be wrong. The figure of merit is the worst-case probability that Bob answers conclusively. We prove that, for every $N\ge3$, no quantum strategy exceeds $1/N$, the value already attained with shared randomness. This holds for finite- and infinite-dimensional entanglement alike, and also in the commuting-operator framework of quantum field theory. In contrast, an explicit family of AQ correlations, one for each $N$, achieves $1/19$, so the ratio between AQ and quantum performance grows without bound. For certificates of equality obtained from binary-outcome boxes, we determine the optimal AQ performance exactly: it exceeds the quantum value precisely when $N\ge9$ and tends to $1/17$ as $N$ grows. The gap persists for longer messages. Entanglement never saves more than one bit over shared randomness, whereas AQ correlations reduce the one-way zero-error cost of deciding whether two $n$-bit strings are equal, with abstention probability at most $1/2$ on every input pair, from $n$ bits to at most $13$.

The quantum bound also yields an exact-zero (face) witness of post-quantumness: a linear inequality that holds for all commuting-operator correlations on a face of the nonsignalling polytope, namely for boxes with prescribed vanishing entries, and that our AQ correlations violate by a factor growing linearly with $N$. For tensor-product models, the inequality underlying this witness, as well as linear lower bounds on entanglement-assisted one-sided-error and zero-error communication for equality, were obtained earlier by Klauck and de Wolf~\cite{KlauckdeWolf2013}; Appendix~\ref{app:kdw} compares their results with ours in detail. Our contributions on the quantum side are the exact one-bit value $1/N$, its refinement for longer messages and the extension of both to commuting operators.

The separation arises exactly at the boundary between the commuting-operator set and the AQ set. These differ only in whether Alice's and Bob's operators commute as operator identities or merely when acting on the shared state. Our quantum bound rests on a sequential-measurement argument that requires Bob's operators to commute with products of Alice's operators. The AQ condition guarantees commutation only for single operators acting on the state, and the correlations we construct exploit precisely this freedom.

The task also separates the AQ set from the next level of the NPA hierarchy. $\mathtt{Level\,1}$, the set $\C_{\mathrm{Vec}}$ of vector correlations, contains the AQ set and coincides with the set singled out by macroscopic locality~\cite{NavascuesWunderlich2010}. Its defining semidefinite constraint involves only the vectors $\psi$, $A_{a|x}\psi$ and $B_{b|y}\psi$, whereas that of the $\mathtt{1+AB}$ level also involves the vectors $A_{a|x}B_{b|y}\psi$. For $\C_{\mathrm{Vec}}$ we determine the optimal certificate of equality from binary-outcome boxes in closed form. It equals $N/[8(N-2)]$ for every $N\ge4$, exceeds the quantum value precisely when $N\ge5$, and decreases to $1/8$, more than twice the AQ limit $1/17$. The proof is elementary: the upper bound follows from the nonnegativity of two squared norms, and an explicit realization in $N+1$ dimensions attains it. Since the level-1 optimum exceeds the AQ optimum for every $N\ge5$, the optimal level-1 boxes form an explicit family in $\C_{\mathrm{Vec}}\setminus\C_\AQ$, and thus separate the two sets analytically in an operational task. Macroscopic locality alone therefore permits certificates of equality approaching $1/8$, and the constraints that distinguish the $\mathtt{1+AB}$ level from $\mathtt{Level\,1}$ lower this limit to $1/17$. Combined with the classical certificate of inequality, level-1 correlations also solve the one-bit task with worst-case probability $N/(10N-16)>1/10$ for every $N\ge4$.

The paper is organized as follows. Section~\ref{sec:models} defines the correlation models and records two closure properties of the AQ set, and Section~\ref{sec:task} introduces the communication task. Section~\ref{sec:quantum} proves the packing lemma, the quantum sum rule and the value $1/N$ for all quantum models. Section~\ref{sec:aq} constructs the AQ boxes $\bm p^\star$, reduces AQ membership on the relevant face of the nonsignalling polytope to a condition on vectors with prescribed inner products, and determines the optimal AQ certificate of equality. Section~\ref{sec:witness} derives the exact-zero witness and determines when $\bm p^\star$ is local, and Section~\ref{sec:kbit} treats longer messages and communication cost. Section~\ref{sec:vec} determines the optimal certificate of equality at $\mathtt{Level\,1}$ of the NPA hierarchy and uses it to separate this level from the AQ set. Section~\ref{sec:discussion} discusses the results and open questions, and Appendix~\ref{app:local} gives an explicit local model for $\bm p^\star$. Appendix~\ref{app:kdw} compares our quantum bounds with the earlier work of Klauck and de Wolf~\cite{KlauckdeWolf2013}. Table~\ref{tab:summary} summarizes the results: it separates the quantities that are determined exactly from those for which only bounds are known, and lists the values of $N$ from which each construction takes effect.

\begin{table}[t]
\centering
\caption{\label{tab:summary}\textbf{Summary of the results.} Upper part: the efficiencies studied, and what is known about each. Lower part: the smallest $N$ from which each explicit construction takes effect. Here $\eta_\Q(N)=1/N$ is the common value of all quantum models.}
\small
\begin{tabular}{>{\raggedright\arraybackslash}p{8.0cm}>{\raggedright\arraybackslash}p{2.9cm}>{\raggedright\arraybackslash}p{3.9cm}}
\toprule
Quantity & Value & Status\\
\midrule
One-bit task, $\Omega$ from $\C_{\mathrm L}$ to $\C_{\mathrm{qc}}$: $\eta_\Omega(N)$ & $1/N$ & exact (Corollary~\ref{cor:quantum})\\
Binary-box AQ certificate of equality: $\eta^{\rm eq}_\AQ(N)$ & Eq.~\eqref{eq:opt_value} & exact (Theorem~\ref{thm:optimal})\\
One-bit task, almost-quantum: $\eta_\AQ(N)$ & $\ge1/19$; Eq.~\eqref{eq:opt_twosided} & lower bounds only; optimum open\\
Binary-box certificate of equality, vector correlations: $\eta^{\rm eq}_{\C_{\mathrm{Vec}}}(N)$ & Eq.~\eqref{eq:vec_main} & exact (Proposition~\ref{prop:vec})\\
One-bit task, vector correlations: $\eta_{\C_{\mathrm{Vec}}}(N)$ & $\ge N/(10N-16)$ & lower bound only\\
\midrule
Construction & First $N$ & Reference\\
\midrule
Optimal vector-correlation certificate exceeds $1/N$ & $5$ & Proposition~\ref{prop:vec}\\
Optimal vector-correlation box lies outside $\C_\AQ$ & $5$ & Section~\ref{sec:vec}\\
Optimal AQ certificate of equality exceeds $1/N$ & $9$ & Theorem~\ref{thm:optimal}\\
Mixture with the optimal level-1 boxes exceeds $\eta_\Q(N)$ & $9$ & Eq.~\eqref{eq:opt_twosided}\\
Mixture with the optimal boxes exceeds $\eta_\Q(N)$ & $14$ & Eq.~\eqref{eq:opt_twosided}\\
$\bm p^\star$ is nonlocal & $14$ & Proposition~\ref{prop:local}\\
$\bm p^\star$ violates the face witness~\eqref{eq:witness} & $18$ & Corollary~\ref{cor:witness}\\
$1/19$ protocol exceeds $\eta_\Q(N)$ & $20$ & Theorem~\ref{thm:qstar}\\
\bottomrule
\end{tabular}
\end{table}

\section{Correlation models}\label{sec:models}
Consider two spacelike-separated laboratories that share a source. In each round, Alice chooses a measurement setting $x\in\mathcal X$ and records an outcome $a\in\mathcal A$; Bob independently chooses $y\in\mathcal Y$ and records $b\in\mathcal B$. All four alphabets are finite. Repeating the experiment produces the \emph{behaviour} (or \emph{box}) $\bm p=\{p(a,b|x,y)\}$, with $p(a,b|x,y)\ge0$ and $\sum_{a,b}p(a,b|x,y)=1$. Throughout, $x,y$ denote box settings and $a,b$ box outcomes; the inputs of the communication task are denoted by $s,t$. The data alone do not fix a physical model of the source and the devices, and different assumptions about the two laboratories lead to different sets of admissible behaviours~\cite{Tsirelson1993,Fritz2012,Junge2011,DykemaPaulsenPrakash2017}. The sets relevant here are defined as follows.

Alice's measurements are denoted $\{A_{a|x}\}_a$ and Bob's $\{B_{b|y}\}_b$. In the Hilbert-space models below they are projective, $A_{a|x}=A_{a|x}^\dagger=A_{a|x}^2$ and $\sum_aA_{a|x}=\Id$, and likewise for Bob. For $\C_{\mathrm q}$ and $\C_{\mathrm{qc}}$, allowing general POVMs gives the same sets, by Naimark dilation in the first case and by a less elementary argument in the second~\cite{DykemaPaulsenPrakash2017}.

\smallskip
\noindent$\bm{[\C_{\mathrm{L}}]}$ \emph{Local hidden variables}. Alice's and Bob's outcomes are spacelike separated. Reichenbach's common-cause principle~\cite{Reichenbach1956} then requires that their correlation be explained by a common cause $\lambda$ emitted by the source, and that conditioning on $\lambda$ screen the outcomes off from each other, $p(a,b|x,y,\lambda)=p_A(a|x,\lambda)\,p_B(b|y,\lambda)$. In addition, the settings are assumed to be chosen independently of $\lambda$. Together these yield Bell's local-causality condition~\cite{Bell1964,Bell1990}, $p(a,b|x,y)=\sum_\lambda q(\lambda)\,p_A(a|x,\lambda)\,p_B(b|y,\lambda)$. Quantum violations of this condition admit no classical causal explanation without fine-tuning~\cite{WoodSpekkens2015}.

\smallskip
\noindent$\bm{[\C_{\mathrm{q}}]}$ \emph{Finite-dimensional quantum}. Alice's and Bob's systems have finite-dimensional Hilbert spaces $\mathcal H_A$ and $\mathcal H_B$. The source prepares a unit vector $\psi\in\mathcal H_A\otimes\mathcal H_B$, and $p(a,b|x,y)=\ip{\psi}{(A_{a|x}\otimes B_{b|y})\psi}$. Mixed states give nothing new, since they can be purified in a larger finite-dimensional space.

\smallskip
\noindent$\bm{[\C_{\mathrm{qs}}]}$ \emph{Spatial quantum}. As $\C_{\mathrm q}$, but $\mathcal H_A$ and $\mathcal H_B$ may be infinite dimensional.

\smallskip
\noindent$\bm{[\C_{\mathrm{qa}}]}$ \emph{Approximately quantum}. The closure $\overline{\C_{\mathrm q}}=\overline{\C_{\mathrm{qs}}}$: behaviours that can be approximated to arbitrary precision by finite-dimensional quantum experiments.

\smallskip
\noindent$\bm{[\C_{\mathrm{qc}}]}$ \emph{Commuting-operator quantum}. The two laboratories are not assumed to be tensor factors. Instead there is a single Hilbert space $\mathcal H$, a unit vector $\psi\in\mathcal H$, and Alice's and Bob's observables commute as operators, $[A_{a|x},B_{b|y}]=0$, with $p(a,b|x,y)=\ip{\psi}{A_{a|x}B_{b|y}\psi}$. This is how locality is encoded in algebraic quantum field theory, where observables localized in spacelike-separated regions commute~\cite{HaagKastler1964}.

\smallskip
\noindent$\bm{[\C_\AQ]}$ \emph{Almost quantum}. As $\C_{\mathrm{qc}}$, but commutation is required only on the state prepared by the source:

\begin{definition}[Almost-quantum correlations~\cite{Navascues2015}]\label{def1}
A behaviour $\bm p$ belongs to $\C_\AQ$ if there exist a Hilbert space $\mathcal H$, a unit vector $\psi\in\mathcal H$, and projective measurements $\{A_{a|x}\}_a$, $\{B_{b|y}\}_b$ on $\mathcal H$ such that, for all $a,b,x,y$,
\begin{equation}\label{eq:AQ}
A_{a|x}B_{b|y}\psi=B_{b|y}A_{a|x}\psi\qquad\text{and}\qquad p(a,b|x,y)=\ip{\psi}{A_{a|x}B_{b|y}\psi}.
\end{equation}
\end{definition}

\noindent Thus $\C_\AQ$ differs from $\C_{\mathrm{qc}}$ only in that commutation is imposed on the state $\psi$ rather than as an operator identity. Equivalently, $\C_\AQ$ is $\mathtt{Level\,1+AB}$ of the NPA hierarchy, characterized by a single semidefinite program~\cite{Navascues2015}.

\smallskip
\noindent$\bm{[\C_{\mathrm{Vec}}]}$ \emph{Vector correlations}. There exist a unit vector $\psi$ and vectors $\xi_{a|x},\zeta_{b|y}$ in a real or complex Hilbert space such that $\xi_{a|x}\perp\xi_{a'|x}$ for $a\neq a'$, $\zeta_{b|y}\perp\zeta_{b'|y}$ for $b\neq b'$, $\sum_a\xi_{a|x}=\psi=\sum_b\zeta_{b|y}$, and $p(a,b|x,y)=\ip{\xi_{a|x}}{\zeta_{b|y}}$. This is $\mathtt{Level\,1}$ of the NPA hierarchy~\cite{NPA2007}, and it coincides with the set singled out by macroscopic locality~\cite{NavascuesWunderlich2010}. The authors of Ref.~\cite{DykemaPaulsenPrakash2017} also call this set ``almost quantum''; we follow Ref.~\cite{Navascues2015} and reserve that name for $\C_\AQ$.

\smallskip
\noindent$\bm{[\C_{\mathrm{NS}}]}$ \emph{Nonsignalling}. Neither party's local statistics depend on the other party's setting, as demanded by relativistic causality. That is, $\sum_bp(a,b|x,y)$ does not depend on $y$ and $\sum_ap(a,b|x,y)$ does not depend on $x$; these sums define the marginals $p_A(a|x)$ and $p_B(b|y)$.

\medskip\noindent
For every Bell scenario,
\begin{equation}\label{eq:chain}
\C_{\mathrm L}\subseteq\C_{\mathrm q}\subseteq\C_{\mathrm{qs}}\subseteq\C_{\mathrm{qa}}\subseteq\C_{\mathrm{qc}}\subseteq\C_\AQ\subseteq\C_{\mathrm{Vec}}\subseteq\C_{\mathrm{NS}},
\end{equation}
and these inclusions are known to be strict in some scenario.
The inclusions~\eqref{eq:chain} follow from the definitions. Tensor-product observables $A\otimes\Id$ and $\Id\otimes B$ commute, which gives $\C_{\mathrm{qs}}\subseteq\C_{\mathrm{qc}}$; since $\C_{\mathrm{qc}}$ is closed~\cite{Fritz2012}, also $\C_{\mathrm{qa}}\subseteq\C_{\mathrm{qc}}$. Commutation as an operator identity implies commutation on $\psi$, which gives $\C_{\mathrm{qc}}\subseteq\C_\AQ$. Given an AQ realization, the vectors $\xi_{a|x}=A_{a|x}\psi$ and $\zeta_{b|y}=B_{b|y}\psi$ form a vector realization, so $\C_\AQ\subseteq\C_{\mathrm{Vec}}$. Finally, $\ip{\xi_{a|x}}{\sum_b\zeta_{b|y}}=\|\xi_{a|x}\|^2$ does not depend on $y$, which gives $\C_{\mathrm{Vec}}\subseteq\C_{\mathrm{NS}}$. Each of the following inclusions is strict in some Bell scenario: $\C_{\mathrm L}\subsetneq\C_{\mathrm q}$ by Bell's theorem~\cite{Bell1964}; $\C_{\mathrm q}\subsetneq\C_{\mathrm{qs}}$~\cite{ColadangeloStark2020}; $\C_{\mathrm{qs}}\subsetneq\C_{\mathrm{qa}}$, because $\C_{\mathrm{qs}}$ is not closed~\cite{Slofstra2019}; $\C_{\mathrm{qa}}\subsetneq\C_{\mathrm{qc}}$, a negative answer to Tsirelson's problem~\cite{Tsirelson1993,ScholzWerner2008}, since equality in all scenarios is equivalent to Connes' embedding conjecture~\cite{Junge2011,Fritz2012,Ozawa2013}, which was refuted by $\mathrm{MIP}^*=\mathrm{RE}$~\cite{MIPRE2020}; $\C_{\mathrm{qc}}\subsetneq\C_\AQ$~\cite{Navascues2015}; $\C_\AQ\subsetneq\C_{\mathrm{Vec}}$, for instance by the boxes of Section~\ref{sec:vec}; and $\C_{\mathrm{Vec}}\subsetneq\C_{\mathrm{NS}}$, because the PR box~\cite{PR1994} violates the Tsirelson bound~\cite{Tsirelson1980}, which already holds at NPA $\mathtt{Level\,1}$.

Two closure properties of $\C_\AQ$ are used below.

\begin{lemma}[Closure properties of $\C_\AQ$]\label{lem:closure}
(i)~$\C_\AQ$ is convex and contains $\C_{\mathrm L}$; in particular, mixing AQ boxes by means of shared randomness yields an AQ box. (ii)~If $\bm p_1,\dots,\bm p_r\in\C_\AQ$, then the product box $\bm p(\bm a,\bm b|\bm x,\bm y)=\prod_{k=1}^r p_k(a_k,b_k|x_k,y_k)$, in which each party holds one setting and one outcome for each factor, belongs to $\C_\AQ$.
\end{lemma}

\begin{proof}
(i)~Classical correlations are AQ by Eq.~\eqref{eq:chain}. If $(\psi_k,\{A^{(k)}\},\{B^{(k)}\})$, $k=1,2$, realize two AQ boxes, then the state $\sqrt w\,\psi_1\oplus\sqrt{1-w}\,\psi_2$ with block-diagonal projectors satisfies Eq.~\eqref{eq:AQ} and realizes their mixture with weights $w$ and $1-w$. (ii)~If $(\psi_k,\{A^{(k)}_{a|x}\},\{B^{(k)}_{b|y}\})$, $k=1,\dots,r$, realize AQ boxes, then $\psi=\bigotimes_k\psi_k$ with the product projectors $A_{\bm a|\bm x}=\bigotimes_kA^{(k)}_{a_k|x_k}$ and $B_{\bm b|\bm y}=\bigotimes_kB^{(k)}_{b_k|y_k}$ realizes the product box, and
\[
A_{\bm a|\bm x}B_{\bm b|\bm y}\psi=\bigotimes_kA^{(k)}_{a_k|x_k}B^{(k)}_{b_k|y_k}\psi_k=\bigotimes_kB^{(k)}_{b_k|y_k}A^{(k)}_{a_k|x_k}\psi_k=B_{\bm b|\bm y}A_{\bm a|\bm x}\psi.\qedhere
\]
\end{proof}

\section{Zero-error certification of equality}\label{sec:task}
\begin{figure}[t]
\centering
\includegraphics[width=0.62\linewidth]{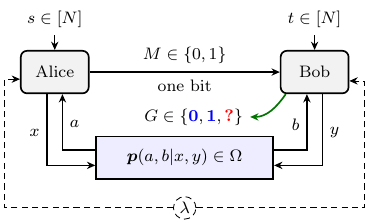}
\caption{\label{fig:task}\textbf{Zero-error certification of equality with one bit.} Alice and Bob receive $s,t\in[N]$ and share a box $\bm p\in\Omega$ together with shared randomness $\lambda$. Alice chooses a setting $x$ depending on $s$, obtains $a$, and sends one bit $M$ to Bob. Bob chooses a setting $y$ depending on $t$ and $M$, obtains $b$, and outputs $G\in\{0,1,\bm ?\}$: $G=1$ certifies equality, $G=0$ certifies inequality and $G=~\bm ?$ denotes abstention. The protocol must never err, and its efficiency is the worst-case probability of a conclusive answer.}
\end{figure}

Alice and Bob each receive one of $N\ge3$ labels, $s,t\in[N]=\{1,\dots,N\}$, and must decide whether they hold the same one; every pair $(s,t)$ is allowed, and the target function is $\EQ_N(s,t)=\Ind\{s=t\}$. Alice may send Bob a single classical bit $M\in\{0,1\}$, after which Bob must give an answer that is never wrong: he either certifies ``equal'' ($G=1$) or ``different'' ($G=0$) with certainty, or abstains ($G=\bm ?$). Abstaining, the analogue of the inconclusive outcome in unambiguous state discrimination~\cite{Ivanovic1987,Dieks1988,Peres1988}, is always safe but uninformative, so the goal is to answer conclusively as often as possible on every input pair, equal and unequal alike (Fig.~\ref{fig:task}).

\begin{definition}[Zero error and efficiency]\label{def2}
A protocol has \emph{zero error} if
\begin{equation}
\Pr\big(G\notin\{\EQ_N(s,t),\bm ?\}\,\big|\,s,t\big)=0\qquad\text{for all } s,t\in[N].
\end{equation}
For a set $\Omega$ of correlations, the optimal \emph{efficiency} is
\begin{equation}\label{eq:eff}
\eta_{\Omega}(N)=\sup_{\text{zero-error protocols over }\Omega}\ \min_{s,t\in[N]}\Pr(G\neq\bm ?\,|\,s,t).
\end{equation}
\end{definition}

\noindent The efficiency is the worst-case probability that Bob answers conclusively, and therefore correctly. We write $\eta_\C\equiv\eta_{\C_{\mathrm L}}$ (shared randomness), $\eta_\Q\equiv\eta_{\C_{\mathrm q}}$ (shared entanglement) and $\eta_\AQ\equiv\eta_{\C_\AQ}$. In the quantum model, the parties share an arbitrary finite-dimensional state $\rho_{AB}$, perform arbitrary local operations and may keep quantum memory until after the message; no quantum system is communicated. To treat all sets of Eq.~\eqref{eq:chain} on an equal footing, we formulate one-way protocols directly in terms of boxes. A protocol over $\Omega$ uses a box $\bm p\in\Omega$ in which Alice's setting is $x=s$ and Bob's setting is the pair $y=(t,m)$; Alice obtains $a$ and sends $m=a$, and Bob's outcome is $G$. All sets in Eq.~\eqref{eq:chain} are convex and closed under local classical post-processing, so shared randomness and local processing can be absorbed into $\bm p$. Because $\bm p$ is nonsignalling, Alice's outcome distribution does not depend on Bob's setting, so letting Bob's setting depend on $m=a$ yields a normalized distribution,
\begin{equation}\label{eq:protocol_box}
\Pr(G|s,t)=\sum_{a}p\big(a,G\,\big|\,s,(t,a)\big).
\end{equation}
For $\Omega=\C_{\mathrm q}$ these are exactly the entanglement-assisted protocols, and for $\Omega=\C_{\mathrm{qc}}$ their commuting-operator counterparts. By construction, $\eta_{\Omega}(N)$ is non-decreasing along the chain~\eqref{eq:chain}.

Equality is the paradigmatic problem of communication complexity, the field introduced by Yao~\cite{Yao1979}. For $n$-bit inputs ($N=2^n$), deterministic protocols require $n+1$ bits (in Yao's model, communication is counted until both parties know the answer), whereas public-coin randomized protocols with bounded error need only $O(1)$ bits~\cite{KushilevitzNisan1997,Newman1991}. Quantum resources reshape these costs in several models, from quantum fingerprinting~\cite{Buhrman2001}, which improves exponentially on the classical simultaneous-message cost of equality~\cite{Ambainis1996,NewmanSzegedy1996}, to the exponential entanglement-assisted saving for the distributed Deutsch--Jozsa problem~\cite{BuhrmanCleveWigderson1998,BuhrmanRMP2010}. Zero-error (Las Vegas) protocols~\cite{MehlhornSchmidt1982,Klauck2000}, which may abstain but never err, show a marked asymmetry: inequality has short certificates, whereas certifying equality requires $n$ bits~\cite{KushilevitzNisan1997}. Prior entanglement does not remove this obstruction: entanglement-assisted one-sided-error and zero-error protocols for equality still require communication linear in $n$, even with interaction~\cite{KlauckdeWolf2013}.
Our task differs from these settings in three respects: the channel is a single noiseless bit and the figure of merit is the worst-case probability of a conclusive answer; errors are forbidden altogether while abstention is unrestricted, with no promise on the inputs; and shared randomness is free, so the classical benchmark is the strongest public-coin model and entire sets of correlations are compared. In this regime the asymmetry noted above reappears sharply: inequality can be certified with probability $1/2$ using shared randomness alone, whereas certifying equality is the bottleneck.

\section{Quantum correlations offer no advantage}\label{sec:quantum}
Our quantum bound rests on the following elementary lemma, a zero-error instance of the sequential-measurement technique underlying the gentle-measurement lemma and quantum union bounds~\cite{Winter1999,Sen2012,Wilde2013,Gao2015}: a test that is certain to fail on a state leaves that state undisturbed~\cite{HeinosaariWolf2010}. Here $T\ge0$ means that $T$ is positive semidefinite, $T\le R$ means $R-T\ge0$, and an \emph{effect} is an operator $0\le F\le\Id$.

\begin{lemma}[Packing]\label{lem:packing}
Let $R$ be a positive trace-class operator, $T_1,\dots,T_N$ positive operators and $F_1,\dots,F_N$ effects such that $T_i\le R$ for all $i$ and $F_jT_i=0$ whenever $j<i$. Then $\sum_i\tr(F_iT_i)\le\tr R$.
\end{lemma}

\begin{proof}
Each effect $F_j$ defines a binary test with outcomes ``yes'' (effect $F_j$) and ``no'' (effect $\Id-F_j$). Apply the tests in the order $F_1,F_2,\dots$ using L\"uders instruments, and stop at the first ``yes''. With $C_1=\Id$ and $C_{i+1}=\sqrt{\Id-F_i}\,C_i$, the outcome ``first yes at test $i$'' has effect $K_i=C_i^\dagger F_iC_i\ge0$. Since $C_{i+1}^\dagger C_{i+1}=C_i^\dagger C_i-K_i$, the sum telescopes, $\sum_iK_i=\Id-C_{N+1}^\dagger C_{N+1}\le\Id$. Let $P_i$ project onto the support of $T_i$. For $j<i$, $F_jT_i=0$ gives $F_jP_i=0$, hence $\sqrt{\Id-F_j}\,P_i=P_i$. Thus $C_iP_i=P_i$ and $C_iT_iC_i^\dagger=T_i$: the earlier tests, being certain to say ``no'' on $T_i$, leave it undisturbed. Therefore $\tr(K_iT_i)=\tr(F_iT_i)$, and since $K_i\ge0$ and $T_i\le R$, $\sum_i\tr(F_iT_i)=\sum_i\tr(K_iT_i)\le\tr[(\sum_iK_i)R]\le\tr R$.
\end{proof}

\noindent Only the zeros below the diagonal are needed, and the labels may be ordered at will; in our applications all off-diagonal products vanish. For tensor-product models a closely related statement is Claim~1 of Ref.~\cite{KlauckdeWolf2013}, discussed in Appendix~\ref{app:kdw}.

\begin{theorem}[Quantum sum rule]\label{thm:sumrule}
Let $N\ge3$. Every zero-error one-bit protocol over $\C_{\mathrm{qc}}$, and hence every entanglement-assisted protocol, satisfies
\begin{equation}\label{eq:sumrule}
\sum_{t=1}^N\Pr(G=1\,|\,t,t)\le1.
\end{equation}
\end{theorem}

\noindent We first outline the proof. The proof proceeds in three steps. Steering Bob's acceptance event onto Alice's side defines, for each message $m$ and input $t$, an unnormalized state $T_{mt}$ of Alice with $T_{mt}\le\rho_A$; zero error forces Alice's encoding effects $E_{m|s}$ to annihilate $T_{mt}$ whenever $s\neq t$. Since $N\ge3$, a third label $u$ shows that the supports associated with the two messages lie in the kernel and in the eigenvalue-one eigenspace of the same effect $E_{0|u}$, so the two message sectors are orthogonal.

Lemma~\ref{lem:packing} bounds each message's contribution by the weight of $\rho_A$ in its corresponding subspace. Orthogonality makes these two weights add to at most one, giving Eq.~\eqref{eq:sumrule}. In the commuting-operator model, the same orthogonality and sequential-measurement argument is carried out directly on the shared Hilbert space.

\paragraph{Proof of Theorem~\ref{thm:sumrule}: tensor-product models.} Let $\rho_{AB}$ be the shared state and $E_{m|s}$ Alice's effect for sending $m$, with $E_{0|s}+E_{1|s}=\Id$. For Bob's input $t$ and received message $m$, let $D_{g|m,t}$ be his effect for announcing $G=g$; thus $D_{1|m,t}$ corresponds to declaring equality. Local processing and quantum memory are included.
\[
\rho_A=\tr_B\rho_{AB},\qquad
T_{mt}=\tr_B\!\left[(\Id\otimes\sqrt{D_{1|m,t}})\rho_{AB}
(\Id\otimes\sqrt{D_{1|m,t}})\right].
\]
Here $T_{mt}$ is Alice's unnormalized state corresponding to Bob's acceptance. Acceptance and rejection together recover $\rho_A$, so $0\le T_{mt}\le\rho_A$. For each message $m$,
\[
\begin{aligned}
\Pr(M=m,G=1\mid s,t)
&=\tr_{AB}\!\left[(E_{m|s}\otimes D_{1|m,t})\rho_{AB}\right]\\
&=\tr_A(E_{m|s}T_{mt}).
\end{aligned}
\]
The first equality is the Born rule; the second follows from the definition of the partial trace and cyclicity of the trace. Summing over the two possible messages gives
\[
\Pr(G=1\mid s,t)=\sum_m\tr(E_{m|s}T_{mt}).
\]
For $s\ne t$, zero error makes this sum zero. Each term is nonnegative, and a positive effect with zero expectation vanishes on the state's support (the subspace it occupies). Hence
\begin{equation}\label{eq:zero}
E_{m|s}T_{mt}=0\qquad(s\ne t).
\end{equation}

The key observation is that the states for the two messages occupy orthogonal subspaces. Take $v\in\supp T_{0t}$ and $w\in\supp T_{1t'}$. Since $N\ge3$, choose $u\notin\{t,t'\}$. Equation~\eqref{eq:zero} gives
\[
E_{0|u}v=0,\qquad E_{1|u}w=0,
\qquad\text{so}\qquad E_{0|u}w=w.
\]
Therefore $\langle v,w\rangle=\langle v,E_{0|u}w\rangle
=\langle E_{0|u}v,w\rangle=0$. Let $\mathcal H_m$ be the closed span of the supports of all $T_{mt}$, and let $\Pi_m$ project onto it. The orthogonality just proved gives $\mathcal H_0\perp\mathcal H_1$. Thus $\Pi_0+\Pi_1$ is the projector onto their combined subspace, so $\Pi_0+\Pi_1\le\Id$.

Fix $m$. Each $T_{mt}$ is supported in $\mathcal H_m$, so projecting onto this subspace leaves it unchanged: $T_{mt}=\Pi_mT_{mt}\Pi_m$. Set $R_m=\Pi_m\rho_A\Pi_m$. Since $\rho_A-T_{mt}\ge0$, we have
\[
R_m-T_{mt}=\Pi_m(\rho_A-T_{mt})\Pi_m\ge0.
\]
Indeed, the expectation of the right-hand operator in any vector $x$ is $\langle\Pi_mx,(\rho_A-T_{mt})\Pi_mx\rangle\ge0$. Hence $0\le T_{mt}\le R_m$ for every $t$: all these states have the same positive upper bound.

We can now apply Lemma~\ref{lem:packing} with $T_t=T_{mt}$, $F_t=E_{m|t}$ and $R=R_m$. The upper-bound condition was just verified; each $E_{m|t}$ is an effect; and Eq.~\eqref{eq:zero} gives $E_{m|t}T_{mt'}=0$ for $t\ne t'$. The lemma does not require these effects, which correspond to different inputs, to sum to the identity. It gives
\[
\sum_t\tr(E_{m|t}T_{mt})
\le\tr R_m
=\tr(\Pi_m\rho_A\Pi_m)
=\tr(\rho_A\Pi_m),
\]
where the last equality uses cyclicity of the trace and $\Pi_m^2=\Pi_m$. By the Born rule, $\tr(\rho_A\Pi_m)$ is the probability that a projection onto $\mathcal H_m$ succeeds on $\rho_A$. This is the weight of $\rho_A$ in that subspace. Adding the two message contributions therefore gives
\[
\begin{aligned}
\sum_t\Pr(G=1\mid t,t)
=\sum_{m=0}^1\sum_t\tr(E_{m|t}T_{mt})\le\tr\!\left[\rho_A(\Pi_0+\Pi_1)\right]\le\tr\rho_A=1.
\end{aligned}
\]
This completes the proof. $\hfil\square$

\paragraph{Proof of Theorem~\ref{thm:sumrule}: commuting-operator models.} Let $\psi$ be the shared unit vector, with $E_{0|s}+E_{1|s}=\Id$ and all Alice effects commuting with all Bob effects. Put $v_{mt}=\sqrt{D_{1|m,t}}\psi$. The Born rule gives
\[
\Pr(G=1\mid s,t)=\sum_m\|\sqrt{E_{m|s}}\,v_{mt}\|^2,
\]
so zero error implies $E_{m|s}v_{mt}=0$ whenever $s\ne t$.

Define
\[
V_{mt}=\{v:E_{m|s}v=0\text{ for every }s\ne t\},
\]
and let $\Pi_m$ project onto the closed span $\mathcal H_m$ of $V_{m1},\ldots,V_{mN}$. Thus $v_{mt}\in\mathcal H_m$. The same third-label argument as above gives $\mathcal H_0\perp\mathcal H_1$, hence $\Pi_0+\Pi_1\le\Id$.

Every Bob effect $D$ preserves each $V_{mt}$, since $E_{m|s}Dv=DE_{m|s}v=0$ for $v\in V_{mt}$ and $s\ne t$. Thus $D$ preserves $\mathcal H_m$ and, being self-adjoint, also $\mathcal H_m^\perp$. Consequently $D\Pi_m=\Pi_mD$.

Now construct one measurement on Alice's side. First measure $\Pi_0$, $\Pi_1$ and $\Id-\Pi_0-\Pi_1$. In branch $m$, apply the sequential L\"uders tests $E_{m|1},\ldots,E_{m|N}$ from Lemma~\ref{lem:packing}. Let $K_{mt}$ be the effect for the first ``yes'' at $t$ in branch $m$. These are outcomes of one measurement, so $\sum_{m,t}K_{mt}\le\Id$; they commute with Bob's effects because all the projectors and update operators used to construct them do.

On $v_{mt}$, the initial projection and all earlier ``no'' outcomes leave the vector unchanged, as in Lemma~\ref{lem:packing}. Therefore
\begin{equation}\label{eq:cqc-packing}
\begin{aligned}
\langle v_{mt},E_{m|t}v_{mt}\rangle
&=\langle v_{mt},K_{mt}v_{mt}\rangle\\
&=\langle\psi,K_{mt}D_{1|m,t}\psi\rangle
\le\langle\psi,K_{mt}\psi\rangle.
\end{aligned}
\end{equation}
The inequality uses commutation and $0\le D_{1|m,t}\le\Id$. Summing over the measurement outcomes gives
\[
\sum_t\Pr(G=1\mid t,t)
\le\sum_{m,t}\langle\psi,K_{mt}\psi\rangle
\le1.
\]
This completes the proof. $\hfil\square$

\begin{corollary}\label{cor:quantum}
For every $N\ge3$, $\eta_{\Omega}(N)=1/N$ for $\Omega\in\{\C_{\mathrm L},\C_{\mathrm q},\C_{\mathrm{qs}},\C_{\mathrm{qa}},\C_{\mathrm{qc}}\}$.
\end{corollary}

\begin{proof}
On a diagonal input $(t,t)$, zero error forces every conclusive answer to be $G=1$, so each term in Eq.~\eqref{eq:sumrule} is at least the worst-case efficiency, which is therefore at most $1/N$ over $\C_{\mathrm{qc}}$, and hence over every smaller set. Conversely, shared randomness attains $1/N$: the parties share a uniformly random label $L\in[N]$, Alice sends $M=\Ind\{s=L\}$, and Bob outputs $G=M$ if $t=L$ and abstains otherwise. Bob answers only when $t=L$, and then $M=\Ind\{s=t\}$, so the protocol never errs and is conclusive with probability exactly $1/N$ on every input pair.
\end{proof}

\noindent Entanglement thus offers no advantage over shared randomness, and the same holds for the closure $\C_{\mathrm{qa}}$, so no sequence of quantum strategies does better either.

\paragraph{Relation to earlier bounds.} For tensor-product models, the inequality underlying Lemma~\ref{lem:packing} and the witness of Section~\ref{sec:witness} appears as Claim~1 in the work of Klauck and de Wolf~\cite{KlauckdeWolf2013}. Combined with the removal of communication by teleportation and guessing, it yields linear lower bounds on entanglement-assisted one-sided-error and zero-error communication for equality, even with interaction. For a single classical bit, the same argument gives $\sum_t\Pr(G=1|t,t)\le2$ and hence $\eta_\Q(N)\le2/N$. Theorem~\ref{thm:sumrule} removes the factor $2$, which is the price of guessing the message, and the resulting value $1/N$ is attained. Appendix~\ref{app:kdw} recalls their results, extends their argument to commuting-operator models (Propositions~\ref{prop:kdw} and~\ref{prop:guess}) and compares the bounds in detail. The almost-quantum constructions and optimizations of Sections~\ref{sec:aq} and~\ref{sec:vec} do not rely on any of these bounds.\par

\section{Almost-quantum certificates of equality}\label{sec:aq}

\subsection{Explicit almost-quantum boxes for every \texorpdfstring{$N$}{N}}\label{sec:qstar}
Consider the box $\bm p^\star$ of Table~\ref{tab:qstar}, with settings $x,y\in[N]$ and binary outcomes $a,b\in\{0,1\}$. It depends only on whether the settings coincide, its marginals are $p_A(0|x)=p_B(0|y)=3/17$, so it is nonsignalling, and it lies on the face
\begin{equation}\label{eq:face}
\mathrm F_0=\{\bm p:\ p(0,0|x,y)=0\ \text{for all}\ x\neq y\}
\end{equation}
of the nonsignalling polytope.

\begin{table}[t]
\centering
\caption{\label{tab:qstar}\textbf{The box $\bm p^\star(a,b|x,y)$.} It depends only on whether the settings coincide and is almost quantum for every $N$.}
\begin{tabular}{lcccc}
\toprule
 & $p(0,0)$ & $p(0,1)$ & $p(1,0)$ & $p(1,1)$\\
\midrule
$x=y$ & $1/17$ & $2/17$ & $2/17$ & $12/17$\\
$x\neq y$ & $0$ & $3/17$ & $3/17$ & $11/17$\\
\bottomrule
\end{tabular}
\end{table}

\begin{theorem}\label{thm:qstar}
For every $N\ge2$, $\bm p^\star\in\C_\AQ$. Consequently, $\eta_\AQ(N)\ge1/19$ for every $N\ge3$, and $\eta_\AQ(N)/\eta_\Q(N)\ge N/19$.
\end{theorem}

\noindent Membership in $\C_\AQ$ is proved in Section~\ref{sec:gram} by an explicit realization: the AQ state and all projectors are built from $3N+1$ real vectors in $\mathbb R^{2N+3}$, and membership is verified by direct multiplication, without numerical semidefinite programming. The dimension $2N+3$ is a feature of this mathematical representation, not a physical implementation cost. The protocol has three ingredients.

\emph{(a) Certificate of equality.} Alice uses the setting $x=s$, obtains $a$ and sends $M=a$. Bob uses the setting $y=t$, obtains $b$, outputs $G=1$ if $M=b=0$ and abstains otherwise. By Table~\ref{tab:qstar}, $\Pr(G=1|s=t)=1/17$ and $\Pr(G=1|s\neq t)=0$. Bob's setting does not depend on $M$, so the box is used as a plain black box.

\emph{(b) Classical certificate of inequality.} The parties share independent fair bits $Z_1,\dots,Z_N$. Alice sends $M=Z_s$; Bob outputs $G=0$ if $M\neq Z_t$ and abstains otherwise. Then $\Pr(G=0|s=t)=0$ and $\Pr(G=0|s\neq t)=1/2$.

\emph{(c) Mixing.} Using shared randomness, the parties run (a) with probability $17/19$ and (b) with probability $2/19$. Neither branch errs, and the probability of a conclusive answer is $17/19\times1/17=1/19$ for $s=t$ and $2/19\times1/2=1/19$ for $s\neq t$. Mixing with shared randomness keeps the resource almost quantum, because $\C_\AQ$ is convex and contains all classical correlations (Lemma~\ref{lem:closure}). Together with Corollary~\ref{cor:quantum}, this gives an efficiency ratio of at least $N/19$, which grows without bound.

\subsection{Reduction to vectors with prescribed inner products}\label{sec:gram}
The AQ condition~\eqref{eq:AQ} turns AQ membership of boxes in $\mathrm F_0$ into a problem about $3N+1$ vectors. For binary outcomes write $A_x\equiv A_{0|x}$ and $B_y\equiv B_{0|y}$.

\begin{lemma}[Gram reduction]\label{lem:gram}
Let $\bm p\in\mathrm F_0$ have binary outcomes, and write $p_A=p_A(0|x)$, $p_B=p_B(0|y)$ and $q=p(0,0|x,x)$. Then $\bm p\in\C_\AQ$ if and only if there exist vectors $\psi,a_x,b_x,w_x$ ($x\in[N]$) in a real Hilbert space with $\|\psi\|=1$ and
\begin{equation}\label{eq:opt_gram}
\begin{aligned}
&\ip{\psi}{a_x}=\|a_x\|^2=p_A,\qquad \ip{\psi}{b_x}=\|b_x\|^2=p_B,\\
&\ip{\psi}{w_x}=\|w_x\|^2=\ip{a_x}{w_x}=\ip{b_x}{w_x}=\ip{a_x}{b_x}=q,\\
&\ip{a_x}{b_y}=\ip{a_x}{w_y}=\ip{b_x}{w_y}=0\qquad(x\neq y).
\end{aligned}
\end{equation}
For clarity we state the lemma for $x$-independent $p_A,p_B,q$; the general case is identical with $x$-dependent values.
\end{lemma}

\begin{proof}
\emph{Necessity.} Given an AQ realization, set $a_x=A_x\psi$, $b_x=B_x\psi$ and $w_x=A_xB_x\psi=B_xA_x\psi$. The first two lines of Eq.~\eqref{eq:opt_gram} follow from $A_x^2=A_x$, $B_x^2=B_x$ and commutation on $\psi$. For $x\neq y$, $\|A_xB_y\psi\|^2=\ip{\psi}{B_yA_xB_y\psi}=\ip{\psi}{A_xB_y\psi}=p(0,0|x,y)=0$, so $A_xB_y\psi=B_yA_x\psi=0$. Hence $\ip{a_x}{b_y}=0$, $\ip{a_x}{w_y}=\ip{A_xB_y\psi}{A_y\psi}=0$, and similarly $\ip{b_x}{w_y}=0$. All prescribed inner products are real, so taking the real part of the Gram matrix yields real vectors with the same properties.

\noindent
\emph{Sufficiency.} Let $\mathcal H$ be the real Hilbert space spanned by $\psi$ and the vectors $a_x,b_x,w_x$ (its complexification may be used instead). For a subspace $V\subseteq\mathcal H$, let $\Pi_V$ denote the projector onto $V$, i.e.\ $\Pi_V^2=\Pi_V=\Pi_V^\dagger$ and $\Pi_V\mathcal H=V$; recall that $\Pi_Vv=u$ whenever $u\in V$ and $v-u\perp V$. For each $x\in[N]$ define
\begin{align*}
A_{0|x}\equiv A_x=\Pi_{\spn\{a_x,w_x\}},\quad A_{1|x}=\Id-A_x,\qquad
B_{0|x}\equiv B_x=\Pi_{\spn\{b_x,w_x\}},\quad B_{1|x}=\Id-B_x,
\end{align*}
so that $\{A_{a|x}\}_a$ and $\{B_{b|y}\}_b$ are projective measurements. Let $x\neq y$. By Eq.~\eqref{eq:opt_gram}, each of $\psi-a_x$, $b_x-w_x$ and $b_y$ is orthogonal to both $a_x$ and $w_x$, and each of $\psi-b_x$, $a_x-w_x$ and $a_y$ is orthogonal to both $b_x$ and $w_x$. Hence
\begin{align*}
A_x\psi=a_x,\quad A_xb_x=w_x,\quad A_xb_y=0,\qquad
B_x\psi=b_x,\quad B_xa_x=w_x,\quad B_xa_y=0,
\end{align*}
and therefore, for all $x,y\in[N]$, $A_xB_y\psi=A_xb_y=\delta_{xy}w_x=B_ya_x=B_yA_x\psi$. Since $A_{1|x}=\Id-A_x$ and $B_{1|y}=\Id-B_y$, the identity $A_{a|x}B_{b|y}\psi=B_{b|y}A_{a|x}\psi$ extends from $(a,b)=(0,0)$ to all outcome pairs; for example, $A_{1|x}B_y\psi=b_y-\delta_{xy}w_x=B_yA_{1|x}\psi$. Thus Eq.~\eqref{eq:AQ} holds. The realized box has $p(0,0|x,y)=\ip{\psi}{A_xB_y\psi}=\delta_{xy}\ip{\psi}{w_x}=\delta_{xy}q$, $p_A(0|x)=\ip{\psi}{a_x}=p_A$ and $p_B(0|y)=\ip{\psi}{b_y}=p_B$. For binary outcomes these values fix all entries, so the realized box is $\bm p$.
\end{proof}

\paragraph{Proof of Theorem~\ref{thm:qstar}.}
Let $\{\psi,e_1,e_2\}\cup\{f_x,g_x\}_{x\in[N]}$ be an orthonormal basis of $\mathbb R^{2N+3}$ and define
\begin{equation}\label{eq:opt_vectors}
\begin{aligned}
a_x&=\tfrac{3}{17}\psi+\tfrac{1}{34}\big(\sqrt{51}\,e_1+\sqrt{15}\,e_2\big)+\tfrac{1}{2\sqrt{17}}\big(f_x+\sqrt5\,g_x\big),\\
b_x&=\tfrac{3}{17}\psi+\tfrac{1}{34}\big(-\sqrt{51}\,e_1+\sqrt{15}\,e_2\big)+\tfrac{1}{2\sqrt{17}}\big(-f_x+\sqrt5\,g_x\big),\\
w_x&=\tfrac{1}{17}\psi-\tfrac{2\sqrt{15}}{85}\,e_2+\tfrac{2}{\sqrt{85}}\,g_x.
\end{aligned}
\end{equation}
Direct multiplication shows that these vectors satisfy Eq.~\eqref{eq:opt_gram} with $p_A=p_B=3/17$ and $q=1/17$, and that $\ip{a_x}{a_y}=\ip{b_x}{b_y}=3/34$ and $\ip{w_x}{w_y}=1/85$ for $x\neq y$. By Lemma~\ref{lem:gram}, the projectors $A_x=\Pi_{\spn\{a_x,w_x\}}$ and $B_y=\Pi_{\spn\{b_y,w_y\}}$ realize a box in $\C_\AQ$ with $p(0,0|x,y)=\delta_{xy}/17$ and $p_A(0|x)=p_B(0|y)=3/17$; the remaining entries of Table~\ref{tab:qstar} follow from normalization. Since only the equality or inequality of labels enters, the construction is valid for every $N\ge2$. By Lemma~\ref{lem:closure}(i), mixing with shared randomness preserves membership in $\C_\AQ$, so the protocol of Section~\ref{sec:qstar} uses AQ resources and attains $1/19$.\hfill$\square$

\subsection{Optimal almost-quantum certification of equality}\label{sec:optimal}
The box $\bm p^\star$ is not an isolated construction: it is the large-$N$ limit of the optimal AQ boxes. To state this precisely we isolate the hard direction of the task. For binary-outcome boxes, the protocol of (a) never errs if and only if $\bm p\in\mathrm F_0$, and its worst-case probability of certifying equality is $\min_xp(0,0|x,x)$. We therefore define
\begin{equation}\label{eq:opt_def}
\eta^{\rm eq}_{\Omega}(N)=\sup_{\bm p\in\Omega\cap\mathrm F_0}\ \min_x\,p(0,0|x,x).
\end{equation}
By Theorem~\ref{thm:sumrule}, $\eta^{\rm eq}_{\Omega}(N)=1/N$ for every $\Omega$ from $\C_{\mathrm L}$ to $\C_{\mathrm{qc}}$. Mixing an equality certificate of efficiency $\eta^{\rm eq}$ with the classical inequality certificate, as in (c), gives
\begin{equation}\label{eq:opt_twosided}
\eta_{\Omega}(N)\ \ge\ \frac{\eta^{\rm eq}_{\Omega}(N)}{1+2\,\eta^{\rm eq}_{\Omega}(N)}.
\end{equation}

\begin{theorem}[Optimal AQ equality certification]\label{thm:optimal}
For every $N\ge3$,
\begin{equation}\label{eq:opt_value}
\eta^{\rm eq}_\AQ(N)=
\begin{cases}
1/N, & 3\le N\le 8,\\[4pt]
N^2/(17N^2-80N+64), & N\ge 8.
\end{cases}
\end{equation}
In particular, for $N\ge8$,
\begin{equation}\label{eq:opt_gap}
\eta^{\rm eq}_\AQ(N)-\eta^{\rm eq}_\Q(N)=\frac{(N-1)(N-8)^2}{N\,(17N^2-80N+64)},
\end{equation}
so AQ correlations outperform quantum ones if and only if $N\ge9$, and $\eta^{\rm eq}_\AQ(N)$ decreases monotonically to $1/17$ as $N\to\infty$.
\end{theorem}

\begin{figure}[t]
\centering
\includegraphics[width=1\linewidth]{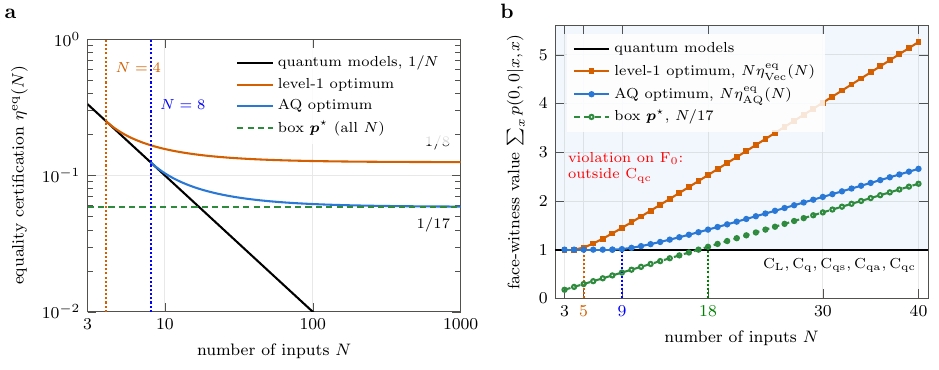}
\caption{\label{fig:efficiency}\textbf{Certifying equality with almost-quantum and level-1 correlations.} \textbf{(a)}~Optimal probability of certifying equality with binary-outcome boxes. For every quantum model, from finite-dimensional entanglement to commuting operators, the optimum is $1/N$, the value attained with shared randomness (black). The optimal almost-quantum value of Theorem~\ref{thm:optimal} (blue) coincides with $1/N$ for $N\le8$, exceeds it for every $N\ge9$, and tends to $1/17$. The boxes $\bm p^\star$ attain $1/17$ for every $N$ (dashed, green). The optimum over $\mathtt{Level\,1}$ of the NPA hierarchy, $\eta^{\rm eq}_{\C_{\mathrm{Vec}}}(N)=N/[8(N-2)]$ (Proposition~\ref{prop:vec}; orange), coincides with $1/N$ for $N\le4$, exceeds it for every $N\ge5$, and tends to $1/8$. It exceeds the AQ optimum for every $N\ge5$, so the boxes that attain it are not almost quantum. \textbf{(b)}~Value of the exact-zero (face) witness $\sum_xp(0,0|x,x)$ on the face $\mathrm F_0$. Every box in $\C_{\mathrm{qc}}\cap\mathrm F_0$, and hence every quantum box on this face, satisfies $\sum_xp(0,0|x,x)\le1$ (black, Corollary~\ref{cor:witness}); values in the shaded region are post-quantum for boxes lying exactly on $\mathrm F_0$. The optimal almost-quantum boxes (blue) give $N\eta^{\rm eq}_\AQ(N)$, which exceeds $1$ for every $N\ge9$. The box $\bm p^\star$ (green) gives $N/17$, which exceeds $1$ for every $N\ge18$. The optimal level-1 boxes (orange) give $N\eta^{\rm eq}_{\C_{\mathrm{Vec}}}(N)=N^2/[8(N-2)]$, which exceeds $1$ for every $N\ge5$. Dotted lines and coloured labels on the horizontal axis mark the first violations.}
\end{figure}

\noindent The proof, given below, has two parts. First, the AQ condition~\eqref{eq:AQ} reduces membership of a box in $\mathrm F_0$ to the existence of $3N+1$ vectors with prescribed inner products, and permutation symmetry reduces the optimization to six real parameters subject to two small linear matrix inequalities. Second, an explicit dual certificate gives the upper bound and an explicit primal solution attains it. Figure~\ref{fig:efficiency} compares the result with the quantum value and with the optimum over $\mathtt{Level\,1}$ of the NPA hierarchy, derived in Section~\ref{sec:vec}. The boxes $\bm p^\star$ attain $1/17$ for every $N$, and no construction valid for all $N$ at once can do better (see the end of the proof below). With the $N$-dependent optimal boxes, Eq.~\eqref{eq:opt_twosided} gives $\eta_\AQ(N)\ge N^2/(19N^2-80N+64)$ for $N\ge8$, which exceeds the quantum value $1/N$ for every $N\ge14$. Theorem~\ref{thm:optimal} concerns binary-outcome boxes in which Bob accepts on a single outcome; whether more general boxes, or settings that depend on the message, can do better is left open.

\paragraph{Symmetry reduction to two small matrix inequalities.} By Lemma~\ref{lem:gram}, computing $\eta^{\rm eq}_\AQ(N)$ in Eq.~\eqref{eq:opt_def} is a semidefinite program. A real symmetric matrix is positive semidefinite if and only if it is the Gram matrix of real vectors. So we maximize $\min_x q_x$ over positive semidefinite matrices $\Gamma$, with rows and columns indexed by $\psi$ and $a_x,b_x,w_x$ ($x\in[N]$), whose entries satisfy Eq.~\eqref{eq:opt_gram} with label-dependent values $p_{A,x}$, $p_{B,x}$ and $q_x$. Here $M\succeq0$ means that the real symmetric matrix $M$ is positive semidefinite. Equation~\eqref{eq:opt_gram} fixes every entry of $\Gamma$ except the inner products between vectors of the same type at different labels, $\ip{a_x}{a_y}$, $\ip{b_x}{b_y}$ and $\ip{w_x}{w_y}$ for $x\neq y$. These are free variables.

\emph{Step 1 (symmetrization):} Relabelling $x\mapsto\pi(x)$, $\pi\in S_N$, maps feasible matrices to feasible matrices and leaves the objective unchanged. The feasible set is convex and $\min_xq_x$ is concave. Hence the average $\bar\Gamma=\frac{1}{N!}\sum_{\pi\in S_N}\Gamma^{\pi}$ of the relabelled copies of a feasible $\Gamma$ is feasible and has an objective value at least as large. The entries of $\bar\Gamma$ depend only on whether two labels coincide. We may therefore assume, without loss of generality, that
\begin{align*}
\ip{\psi}{a_x}=p_A,~~ \ip{\psi}{b_x}=p_B,~~ \ip{\psi}{w_x}=q,~~
\ip{a_x}{a_y}=\alpha,~~ \ip{b_x}{b_y}=\beta,~~ \ip{w_x}{w_y}=\omega\quad(x\neq y).
\end{align*}
This leaves six real parameters $p_A,p_B,q,\alpha,\beta,\omega$, and the objective becomes $q$.

\emph{Step 2 (block structure):} Order the vectors as $\psi$, $(a_1,b_1,w_1),\dots,(a_N,b_N,w_N)$, and set $r=(p_A,p_B,q)^{\mathsf T}$,
\begin{equation}
D=\begin{pmatrix}p_A&q&q\\ q&p_B&q\\ q&q&q\end{pmatrix},\qquad
E=\begin{pmatrix}\alpha&0&0\\0&\beta&0\\0&0&\omega\end{pmatrix}.
\end{equation}
Here $D$ collects the inner products within one label, and $E$ those between different labels. $E$ is diagonal because Eq.~\eqref{eq:opt_gram} sets $\ip{a_x}{b_y}=\ip{a_x}{w_y}=\ip{b_x}{w_y}=0$ for $x\neq y$. For illustration, take $N=2$. With rows and columns ordered as $\psi,a_1,b_1,w_1,a_2,b_2,w_2$, Eq.~\eqref{eq:opt_gram} and Step~1 give
\begin{align*}
\Gamma\big|_{N=2}\;=\;
\begin{array}{c|c|ccc|ccc}
 & \psi & a_1 & b_1 & w_1 & a_2 & b_2 & w_2\\ \hline
\psi & 1 & p_A & p_B & q & p_A & p_B & q\\ \hline
a_1 & p_A & p_A & q & q & \alpha & 0 & 0\\
b_1 & p_B & q & p_B & q & 0 & \beta & 0\\
w_1 & q & q & q & q & 0 & 0 & \omega\\ \hline
a_2 & p_A & \alpha & 0 & 0 & p_A & q & q\\
b_2 & p_B & 0 & \beta & 0 & q & p_B & q\\
w_2 & q & 0 & 0 & \omega & q & q & q
\end{array}
\end{align*}
The first row and column consist of $1$ and copies of $r$, every diagonal $3\times3$ block equals $D$, and every off-diagonal block equals $E$. The same pattern holds for every $N$:
\begin{align*}
\Gamma&=
\begin{array}{c|c|cccc}
 & \psi & (a_1,b_1,w_1) & (a_2,b_2,w_2) & \cdots & (a_N,b_N,w_N)\\ \hline
\psi & 1 & r^{\mathsf T} & r^{\mathsf T} & \cdots & r^{\mathsf T}\\ \hline
(a_1,b_1,w_1) & r & D & E & \cdots & E\\
(a_2,b_2,w_2) & r & E & D & \cdots & E\\
\vdots & \vdots & \vdots & \vdots & \ddots & \vdots\\
(a_N,b_N,w_N) & r & E & E & \cdots & D
\end{array}\\
&=\begin{pmatrix}1&\bm1^{\mathsf T}\otimes r^{\mathsf T}\\ \bm1\otimes r&\Id_N\otimes D+(J_N-\Id_N)\otimes E\end{pmatrix},
\end{align*}
where $\bm1\in\mathbb R^N$ is the all-ones vector and $J_N=\bm1\bm1^{\mathsf T}$.

\emph{Step 3 (block diagonalization):} Let $u_0=\bm1/\sqrt N,u_1,\dots,u_{N-1}$ be an orthonormal basis of $\mathbb R^N$, and let $U$ be the matrix with columns $u_k$. Replace the label vectors by the combinations
\[
\hat a_k=\sum_{x}(u_k)_x\,a_x,\qquad \hat b_k=\sum_{x}(u_k)_x\,b_x,\qquad \hat w_k=\sum_{x}(u_k)_x\,w_x\qquad(k=0,\dots,N-1),
\]
which is the orthogonal change of basis $1\oplus(U\otimes\Id_3)$ applied to $\Gamma$; in particular, $\hat a_0=N^{-1/2}\sum_xa_x$ is the label average. For $v,v'\in\{a,b,w\}$, the entries of $\Gamma$ are $\ip{v_x}{v'_y}=\delta_{xy}(D-E)_{vv'}+E_{vv'}$ and $\ip{\psi}{v_x}=r_v$. Using $\sum_x(u_k)_x(u_l)_x=\delta_{kl}$ and $\sum_x(u_k)_x=\bm1^{\mathsf T}u_k=\sqrt N\,\delta_{k0}$, we obtain
\[
\ip{\psi}{\hat v_k}=\sqrt N\,\delta_{k0}\,r_v,\qquad
\ip{\hat v_k}{\hat v'_l}=\delta_{kl}\,(D-E)_{vv'}+N\,\delta_{k0}\delta_{l0}\,E_{vv'}.
\]
For $N=2$, for instance, $u_0=(1,1)/\sqrt2$ and $u_1=(1,-1)/\sqrt2$, so that $\hat a_0=(a_1+a_2)/\sqrt2$ is the label average and $\hat a_1=(a_1-a_2)/\sqrt2$ the label difference, and similarly for $b$ and $w$. In the new basis $\Gamma$ takes the block form
\begin{align*}
\Gamma\;\cong\;
\begin{array}{c|cc|ccc}
 & \psi & (\hat a_0,\hat b_0,\hat w_0) & (\hat a_1,\hat b_1,\hat w_1) & \cdots & (\hat a_{N-1},\hat b_{N-1},\hat w_{N-1})\\ \hline
\psi & 1 & \sqrt N\,r^{\mathsf T} & 0 & \cdots & 0\\
(\hat a_0,\hat b_0,\hat w_0) & \sqrt N\,r & D+(N-1)E & 0 & \cdots & 0\\ \hline
(\hat a_1,\hat b_1,\hat w_1) & 0 & 0 & D-E & \cdots & 0\\
\vdots & \vdots & \vdots & \vdots & \ddots & \vdots\\
(\hat a_{N-1},\hat b_{N-1},\hat w_{N-1}) & 0 & 0 & 0 & \cdots & D-E
\end{array}
\end{align*}
That is, $\Gamma\;\cong\;L\oplus\underbrace{(D-E)\oplus\cdots\oplus(D-E)}_{N-1\ \text{copies}}$ with
\begin{align*}
L=\begin{pmatrix}1&\sqrt N\,r^{\mathsf T}\\ \sqrt N\,r&D+(N-1)E\end{pmatrix}
=\begin{pmatrix}
1&\sqrt N\,p_A&\sqrt N\,p_B&\sqrt N\,q\\
\sqrt N\,p_A&p_A+(N-1)\alpha&q&q\\
\sqrt N\,p_B&q&p_B+(N-1)\beta&q\\
\sqrt N\,q&q&q&q+(N-1)\omega
\end{pmatrix}
\end{align*}
and
\begin{align*}
D-E=\begin{pmatrix}p_A-\alpha&q&q\\ q&p_B-\beta&q\\ q&q&q-\omega\end{pmatrix}.
\end{align*}
An orthogonal change of basis preserves positive semidefiniteness, and a block-diagonal matrix is positive semidefinite if and only if each block is. Hence, for $N\ge2$, $\Gamma\succeq0$ if and only if
\begin{equation}\label{eq:opt_lmi}
M_1\equiv D-E\succeq0\qquad\text{and}\qquad L\succeq0.
\end{equation}
Consequently,
\[
\eta^{\rm eq}_\AQ(N)=\max\bigl\{\,q:\ (p_A,p_B,q,\alpha,\beta,\omega)\in\mathbb R^6\ \text{satisfy Eq.~\eqref{eq:opt_lmi}}\bigr\},
\]
a semidefinite program in six real variables with one $3\times3$ and one $4\times4$ matrix inequality. The feasible set is compact, so the maximum is attained.

\paragraph{Proof of Theorem~\ref{thm:optimal}.} By the symmetry reduction above, $\eta^{\rm eq}_\AQ(N)$ is the optimal value of the semidefinite program
\begin{equation}\label{eq:opt_sdp}
\text{maximize } q\quad\text{over }(p_A,p_B,q,\alpha,\beta,\omega)\in\mathbb R^6\quad\text{subject to}\quad M_1\succeq0,\ \ L\succeq0 .
\end{equation}
We prove matching bounds: an upper bound from an explicit dual certificate, and a lower bound from explicit feasible points. Throughout, $e_a,e_b,e_w$ denote the standard basis of $\mathbb R^3$, ordered as the rows of $D$ and $E$, and $\Delta_N=17N^2-80N+64$. Since $L_{11}=1>0$, $L\succeq0$ if and only if its Schur complement
\[
S\equiv D+(N-1)E-N\,rr^{\mathsf T}=(D-E)+N\,(E-rr^{\mathsf T})
\]
is positive semidefinite; we use this form to check feasibility.

\noindent$\bullet$ \emph{Upper bound: the principle.} Let $Y$ be a positive semidefinite $3\times3$ matrix and $Z$ a positive semidefinite $4\times4$ matrix. The trace of a product of two positive semidefinite matrices is nonnegative, so every feasible point of~\eqref{eq:opt_sdp} satisfies
\[
\tr(YM_1)+\tr(ZL)\ge0 .
\]
The left-hand side is an affine function of the six parameters. Suppose $Y$ and $Z$ are chosen so that the coefficients of $p_A,p_B,\alpha,\beta,\omega$ vanish and the constant term equals $1$. Then the left-hand side equals $1-c\,q$ for some number $c$, and the inequality gives $q\le1/c$ whenever $c>0$.

\noindent$\bullet$ \emph{Upper bound: the certificate.} For $t\in[0,2]$, let
\[
\rho_t=-\frac{N+2(N-1)t-t^2}{N},\quad
y_t=(t,\,t,\,\rho_t),\quad
z_t=\Big(1,-\tfrac{t^2}{2\sqrt N},-\tfrac{t^2}{2\sqrt N},\tfrac{\rho_t}{\sqrt N}\Big),\quad
\bm b=(0,1,-1,0),
\]
and
\[
Y_t=\tfrac{N-1}{N}\,y_ty_t^{\mathsf T},\qquad
Z_t=z_tz_t^{\mathsf T}+\mu_t\,\bm b\bm b^{\mathsf T},\qquad
\mu_t=\frac{t^2(4-t^2)}{4N}.
\]
Both matrices are positive semidefinite, since $vv^{\mathsf T}\succeq0$ for every vector $v$ and $\mu_t\ge0$ for $t\le2$. Expanding the traces, one finds that the coefficients of $p_A,p_B,\alpha,\beta,\omega$ vanish identically in $t$, that the constant term is $(Z_t)_{11}=1$, and that
\begin{equation}\label{eq:opt_dual}
\tr(Y_tM_1)+\tr(Z_tL)=1-c_t\,q,\qquad
c_t=1+\frac{N-1}{N^2}\,h(t),\qquad
h(t)=N^2-\big(N-2t-t^2\big)^2 .
\end{equation}
Hence $q\le1/c_t$ for every $t\in[0,2]$ with $c_t>0$. The certificate $(Y_t,Z_t)$ is a feasible point of the dual program; how it was found plays no role in the proof, which needs only the identity~\eqref{eq:opt_dual}.

\noindent$\bullet$ \emph{Upper bound: the best $t$.} The bound $1/c_t$ is smallest when $h(t)$ is largest. By Eq.~\eqref{eq:opt_dual}, $h(t)\le N^2$, with equality if and only if $t^2+2t=N$, i.e., $t=t_0\equiv\sqrt{N+1}-1$. Moreover $t_0\le2$ if and only if $N\le8$.
\begin{itemize}
\item For $3\le N\le8$, take $t=t_0\in[0,2]$. Then $h(t_0)=N^2$, so $c_{t_0}=N$ and $q\le1/N$.
\item For $N\ge8$, $N-2t-t^2\ge N-8\ge0$ on $[0,2]$ and decreases with $t$, so $h$ is largest at $t=2$, where $h(2)=N^2-(N-8)^2=16(N-4)$. Then $c_2=1+16(N-1)(N-4)/N^2=\Delta_N/N^2>0$, so $q\le N^2/\Delta_N$.
\end{itemize}

\noindent$\bullet$ \emph{Lower bound for $3\le N\le8$.} The parties share a uniformly random label $u\in[N]$, and each outputs $0$ exactly when their setting equals $u$. Outcome $(0,0)$ then occurs only when $x=y=u$, so this local box lies in $\mathrm F_0$, and $p(0,0|x,x)=\Pr(u=x)=1/N$. Local boxes are almost quantum, so $\eta^{\rm eq}_\AQ(N)\ge1/N$.

\noindent$\bullet$ \emph{Lower bound for $N\ge8$.} Take
\begin{equation}\label{eq:opt_primal}
q_N=\frac{N^2}{\Delta_N},~
p_A=p_B=\frac{3N^2-18N+16}{\Delta_N},~
\alpha=\beta=\frac{3N^2-28N+32}{2\Delta_N},~
\omega=\frac{N^2(N-8)}{(5N-8)\Delta_N}.
\end{equation}
Because $p_A=p_B$ and $\alpha=\beta$, both $M_1$ and $S$ are invariant under exchanging $e_a$ and $e_b$. In the basis $\{(e_a-e_b)/\sqrt2,\,(e_a+e_b)/\sqrt2,\,e_w\}$ they therefore split into a $1\times1$ block and a $2\times2$ block:
\[
M_1\cong\frac{N(N-8)}{2\Delta_N}\ \oplus\ \frac{N}{\Delta_N}\,K_+,\qquad
S\cong\frac{3N(N-1)(N-8)}{2\Delta_N}\ \oplus\ \frac{N(N-1)(N-8)(3N-8)}{\Delta_N^2}\,K_-,
\]
where
\[
K_\pm=\begin{pmatrix}\tfrac{5N-8}{2}&\pm\sqrt2\,N\\[2pt] \pm\sqrt2\,N&\tfrac{4N^2}{5N-8}\end{pmatrix}.
\]
Each $K_\pm$ has positive diagonal entries and determinant $2N^2-2N^2=0$, so it is positive semidefinite (of rank one). For $N\ge8$ all prefactors are nonnegative, so $M_1\succeq0$ and $S\succeq0$, and the point~\eqref{eq:opt_primal} is feasible for~\eqref{eq:opt_sdp}. Lemma~\ref{lem:gram} turns it into an AQ box in $\mathrm F_0$ with $p(0,0|x,x)=q_N$, so $\eta^{\rm eq}_\AQ(N)\ge N^2/\Delta_N$.

\noindent$\bullet$ \emph{Conclusion.} The upper and lower bounds coincide, which proves Eq.~\eqref{eq:opt_value}; at $N=8$ both cases give $1/8$. The value decreases monotonically in $N$: $1/N$ does so for $N\le8$, and for $N\ge8$
\[
\frac{d}{dN}\,\frac{N^2}{\Delta_N}=-\frac{16N(5N-8)}{\Delta_N^2}<0,\qquad \lim_{N\to\infty}\frac{N^2}{\Delta_N}=\frac1{17}.
\]
By Theorem~\ref{thm:sumrule}, the quantum value is $1/N$, and for $N\ge8$
\[
\frac{N^2}{\Delta_N}-\frac1N=\frac{N^3-\Delta_N}{N\Delta_N}=\frac{(N-1)(N-8)^2}{N\Delta_N},
\]
which is Eq.~\eqref{eq:opt_gap}. It is positive if and only if $N\ge9$.

\noindent$\bullet$ \emph{The box $\bm p^\star$.} As $N\to\infty$, the parameters~\eqref{eq:opt_primal} converge to
\[
(p_A,p_B,q)=\big(\tfrac3{17},\tfrac3{17},\tfrac1{17}\big),\qquad \alpha=\beta=\tfrac3{34},\qquad \omega=\tfrac1{85},
\]
the values realized by the vectors~\eqref{eq:opt_vectors}. These limiting values are feasible for every $N$. Indeed, $D-E$ has eigenvalues $0,\,1/34,\,33/170$ and $E-rr^{\mathsf T}$ has eigenvalues $0,\,99/2890,\,3/34$, so $M_1=D-E\succeq0$ and $S=(D-E)+N(E-rr^{\mathsf T})\succeq0$ for all $N$. Thus $\bm p^\star$ attains $1/17$ for every $N$. No construction valid for all $N$ at once can do better: a value $q$ attained for every $N$ satisfies $q\le\eta^{\rm eq}_\AQ(N)$ for every $N$, and since $\eta^{\rm eq}_\AQ(N)$ decreases to $1/17$, $q\le1/17$.\hfill$\square$

\subsection{The separation across correlation models}\label{sec:separation}
Combining Corollary~\ref{cor:quantum}, Theorems~\ref{thm:qstar} and~\ref{thm:optimal}, and Proposition~\ref{prop:vec} of Section~\ref{sec:vec} with the monotonicity of $\eta_{\Omega}$ along the chain~\eqref{eq:chain} locates the separation precisely.

\begin{theorem}\label{thm:models}
For every $N\ge3$:
\begin{enumerate}
\item[(i)] $\eta_{\Omega}(N)=\eta^{\rm eq}_{\Omega}(N)=1/N$ for $\Omega\in\{\C_{\mathrm L},\C_{\mathrm q},\C_{\mathrm{qs}},\C_{\mathrm{qa}},\C_{\mathrm{qc}}\}$;
\item[(ii)] $\eta_\AQ(N)\ge1/19$, while $\eta^{\rm eq}_\AQ(N)$, given by Eq.~\eqref{eq:opt_value}, exceeds $1/N$ if and only if $N\ge9$ and decreases to $1/17$;
\item[(iii)] $\eta^{\rm eq}_{\C_{\mathrm{Vec}}}(3)=1/3$ and $\eta^{\rm eq}_{\C_{\mathrm{Vec}}}(N)=N/[8(N-2)]$ for $N\ge4$, which exceeds $1/N$ if and only if $N\ge5$ and decreases to $1/8$; moreover, $\eta_{\C_{\mathrm{Vec}}}(N)\ge N/(10N-16)>1/10$ for $N\ge4$; and
\item[(iv)] $\eta_{\C_{\mathrm{NS}}}(N)=1$.
\end{enumerate}
In particular, $\eta^{\rm eq}_{\C_{\mathrm{Vec}}}(N)>\eta^{\rm eq}_\AQ(N)$ for every $N\ge5$, and hence $\C_\AQ\subsetneq\C_{\mathrm{Vec}}$.
\end{theorem}

\noindent Parts (i) and (ii) restate Corollary~\ref{cor:quantum}, Theorems~\ref{thm:qstar} and~\ref{thm:optimal}, and the remark after Eq.~\eqref{eq:opt_def}. Part (iii) is Proposition~\ref{prop:vec}, proved in Section~\ref{sec:vec}, combined with Eq.~\eqref{eq:opt_twosided}, whose right-hand side equals $N/(10N-16)$ for $\eta^{\rm eq}=N/[8(N-2)]$. The final statement compares parts (ii) and (iii) (see the end of Section~\ref{sec:vec}); since $\eta^{\rm eq}_{\C_{\mathrm{Vec}}}(N)$ is attained, every box that attains it with $N\ge5$ lies in $\C_{\mathrm{Vec}}\setminus\C_\AQ$. Part (iv) follows from the nonsignalling box with $p(a,b|x,y)=1/2$ if $a\oplus b=\Ind\{x=y\}$ and $0$ otherwise: Alice inputs $x=s$ and sends $m=a$, and Bob inputs $y=t$ and outputs $G=m\oplus b=\Ind\{s=t\}$, which is always conclusive and correct. The unbounded gap therefore opens between the adjacent sets $\C_{\mathrm{qc}}$ and $\C_\AQ$, which differ only in whether Alice's and Bob's observables commute as operators or merely on the state. Consequently, the separation does not depend on the resolution of Tsirelson's problem~\cite{Tsirelson1993,ScholzWerner2008}: neither infinite-dimensional entanglement nor the commuting-operator framework of quantum field theory improves on shared randomness for this task. At the other extreme, general nonsignalling correlations solve the task perfectly, as expected from the triviality of communication complexity with PR-type boxes~\cite{vanDam2013,Brassard2006}. Parts (ii) and (iii) locate a second, bounded separation, between $\C_\AQ$ and $\C_{\mathrm{Vec}}$: the optimal certificates of equality differ for every $N\ge5$, and their limits, $1/17$ and $1/8$, differ by more than a factor of two.

\section{An exact-zero (face) witness of post-quantumness}\label{sec:witness}
Theorem~\ref{thm:sumrule} also yields a simple certificate that a given box is not quantum, provided the box lies exactly on the face $\mathrm F_0$. For a binary-outcome box $\bm p\in\mathrm F_0$, the protocol of (a) is a zero-error one-bit protocol, and Eq.~\eqref{eq:sumrule} becomes a linear inequality on the box.

\begin{corollary}[Exact-zero (face) witness]\label{cor:witness}
For $N\ge3$, every binary-outcome box $\bm p\in\C_{\mathrm{qc}}\cap\mathrm F_0$ satisfies
\begin{equation}\label{eq:witness}
\sum_{x=1}^Np(0,0|x,x)\le1.
\end{equation}
The box $\bm p^\star$ violates Eq.~\eqref{eq:witness} for every $N\ge18$, and the optimal boxes of Theorem~\ref{thm:optimal} violate it for every $N\ge9$. Hence $\{\bm p^\star\}_{N\ge18}$ is an explicit family in $\C_\AQ\setminus\C_{\mathrm{qc}}$, lying outside $\C_{\mathrm q}$, $\C_{\mathrm{qs}}$ and $\C_{\mathrm{qa}}$ as well.
\end{corollary}

\begin{proof}
For $\bm p^\star$ the left-hand side of Eq.~\eqref{eq:witness} equals $N/17$, which exceeds $1$ for $N\ge18$. For the optimal boxes it equals $1$ for $N\le8$ and $N^3/(17N^2-80N+64)$ for $N\ge8$; on the latter branch it exceeds $1$ if and only if $(N-1)(N-8)^2>0$, i.e., for $N\ge9$ (see Fig.~\ref{fig:efficiency}).
\end{proof}

\noindent\emph{Scope of the witness.} Inequality~\eqref{eq:witness} is not valid throughout the quantum set; it holds for boxes whose off-diagonal coincidences $p(0,0|x,y)$, $x\neq y$, vanish exactly. Without this condition it fails already for local boxes. Suppose that with probability $\varepsilon$ both parties output $0$ for every setting, and otherwise both output $1$. Then $p(0,0|x,y)=\varepsilon$ for all $x,y$, while $\sum_xp(0,0|x,x)=N\varepsilon$ exceeds $1$ once $N>1/\varepsilon$. Arbitrarily small off-diagonal coincidences can thus coexist with a large value of the left-hand side, and a violation by a box that lies only approximately on $\mathrm F_0$ certifies nothing. We therefore call Eq.~\eqref{eq:witness} an exact-zero, or face, witness. Neither its growing violation nor the growing efficiency ratio implies noise tolerance or a lower bound on the distance from the quantum set; a quantitative version that allows nonzero off-diagonal coincidences is left for future work. For tensor-product models, Eq.~\eqref{eq:witness} also follows from Claim~1 of Ref.~\cite{KlauckdeWolf2013}, which requires the zeros only for $x>y$ (Appendix~\ref{app:kdw}).

The same triangular-zero bound holds for commuting-operator correlations: apply Alice's L\"uders tests $A_{0|N},\ldots,A_{0|1}$ in decreasing label order. For $v_y=\sqrt{B_{0|y}}\psi$, every earlier test vanishes and leaves $v_y$ unchanged. The sequential-test estimate~\eqref{eq:cqc-packing} in the proof of Theorem~\ref{thm:sumrule}, with the sector projector set to $\Id$, then gives $\sum_y p(0,0|y,y)\le1$. Proposition~\ref{prop:kdw} in Appendix~\ref{app:kdw} gives the argument in full.\par

\noindent For small $N$, however, $\bm p^\star$ is not even nonlocal.

\begin{proposition}[Locality of $\bm p^\star$]\label{prop:local}
Every local box on the face $\mathrm F_0$ satisfies
\begin{equation}\label{eq:localbound}
\sum_{x=1}^N p(0,0|x,x)+\frac1N\sum_{x=1}^N\big[p(0,1|x,x)+p(1,0|x,x)\big]\le1.
\end{equation}
The box $\bm p^\star$ gives $(N+4)/17$ on the left-hand side, and it is local if and only if $N\le13$.
\end{proposition}

\begin{proof}
A local box is a mixture of deterministic strategies $(f,g)$, in which Alice outputs $a=f(x)$ and Bob outputs $b=g(y)$. On $\mathrm F_0$, $p(0,0|x,y)=0$ for $x\neq y$. This is a sum of nonnegative terms, one per strategy, so every strategy with positive weight lies on $\mathrm F_0$ itself. Writing $F_A=f^{-1}(0)$ and $F_B=g^{-1}(0)$ for the inputs on which Alice and Bob output $0$, this means that $F_A\times F_B$ contains no pair $(x,y)$ with $x\neq y$. Hence one of the following holds:
(i)~$F_A=\emptyset$, i.e.\ Alice always outputs $1$;
(ii)~$F_B=\emptyset$, i.e.\ Bob always outputs $1$;
(iii)~$F_A=F_B=\{x_0\}$ for a single label $x_0$.

\noindent It suffices to check each type, since the left-hand side is linear in $\bm p$.
In type (i), only the outcome pair $(1,0)$ can occur, and only for $x\in F_B$, so the left-hand side equals $|F_B|/N\le1$.
Type (ii) gives $|F_A|/N\le1$ in the same way.
In type (iii), the parties output $(0,0)$ on input $x_0$ and $(1,1)$ otherwise, so the left-hand side equals exactly $1$.

For $\bm p^\star$, $p(0,0|x,x)=1/17$ and $p(0,1|x,x)=p(1,0|x,x)=2/17$, so the left-hand side equals $(N+4)/17$. This exceeds $1$ if and only if $N\ge14$.

A local model for $N\le13$ is constructed explicitly in Appendix~\ref{app:local}.
\end{proof}

\noindent By Corollary~\ref{cor:witness}, $\bm p^\star$ lies outside $\C_{\mathrm{qc}}$ for $N\ge18$. Whether it is quantum for $14\le N\le17$ is left open.

\section{Longer messages and communication cost}\label{sec:kbit}
We now let Alice send one of $K$ messages $m\in\{0,\dots,K-1\}$; a $k$-bit message corresponds to $K=2^k$. Let $\eta^{(K)}_{\Omega}(N)$ be the optimal worst-case conclusive probability with $K$ messages and resources $\Omega$, so that $\eta_\Omega=\eta^{(2)}_\Omega$. Longer messages help quantum and classical players almost equally, so the gap to almost-quantum correlations persists.

\emph{Shared randomness.} For $K\le N$ the following protocol attains $(K-1)/N$. The parties share a uniformly random set $\mathcal L\subset[N]$ of $K-1$ labels, listed in a fixed order. Alice sends the position of $s$ in $\mathcal L$ if $s\in\mathcal L$, and the remaining $K$th message otherwise. If $t\in\mathcal L$, Bob outputs $G=1$ when the message is the position of $t$ and $G=0$ otherwise; if $t\notin\mathcal L$, he abstains. Every conclusive answer is correct, and Bob answers whenever $t\in\mathcal L$, which happens with probability $(K-1)/N$ for every input pair. For $N>K$ this is optimal. In a deterministic protocol, Bob can output $G=1$ on $(t,t)$ only if no other label is mapped to the same message as $t$; otherwise he would err on some pair $(s,t)$ with $s\neq t$. Since $N>K$, the $N$ labels cannot be spread one per message, so at most $K-1$ messages have exactly one label mapped to them. Hence $\sum_t\Pr(G=1|t,t)\le K-1$. The same bound holds for mixtures of deterministic protocols, so $\eta^{(K)}_{\C_{\mathrm L}}(N)=(K-1)/N$ for $K<N$; for $K\ge N$, Alice can send $s$ itself and $\eta^{(K)}_{\C_{\mathrm L}}(N)=1$.

\begin{theorem}[Sum rule for $K$ messages]\label{thm:kbit}
Let $N\ge2$ and $K\ge2$. Every zero-error quantum or commuting-operator protocol with $K$ messages satisfies
\begin{equation}\label{eq:kbit}
\sum_{t\neq u}\Pr(G=1\,|\,t,t)\le K-1\qquad\text{for every}\ u\in[N].
\end{equation}
Consequently, $\eta^{(K)}_{\C_{\mathrm{qc}}}(N)\le(K-1)/(N-1)$.
\end{theorem}

\noindent The consequence follows because the left-hand side of Eq.~\eqref{eq:kbit} has $N-1$ terms, each at least $\eta^{(K)}$. The proof adapts that of Theorem~\ref{thm:sumrule}.

With one bit, that argument uses the orthogonality of the two message sectors. With $K$ messages, we instead fix a label $u$. For $t\ne u$, zero error places the support of Alice's steered operator $T_{mt}$, or the conditional vector $\sqrt{D_{1|m,t}}\psi$ in the commuting-operator model, in the kernel of $E_{m|u}$; otherwise Bob would err on $(u,t)$.

The operator inequality $\Pi_{\ker E}\le\Id-E$, valid for every effect $E$, then bounds the total weight of the $K$ message sectors by $\sum_m(\Id-E_{m|u})=(K-1)\Id$. For $K=2$ the bound $1/(N-1)$ is slightly weaker than the bound $1/N$ of Theorem~\ref{thm:sumrule}. Comparing with the classical value $(K-1)/N$, entanglement, even in the commuting-operator framework, improves on shared randomness by at most a factor $N/(N-1)$ for every $K<N$.

\paragraph{Proof of Theorem~\ref{thm:kbit}.} In tensor-product models, define $E_{m|s}$, $D_{1|m,t}$ and $T_{mt}$ as in the proof of Theorem~\ref{thm:sumrule}, now with $m\in\{0,\dots,K-1\}$; zero error again gives Eq.~\eqref{eq:zero}.
 Fix $u$, and let $\Pi_m$ project onto the closed linear span $\mathcal W_m=\overline{\spn}\{\supp T_{mt}:t\neq u\}$. By Eq.~\eqref{eq:zero}, $\mathcal W_m$ lies in the kernel of $E_{m|u}$, so $\Pi_m\le\Pi_{\ker E_{m|u}}\le\Id-E_{m|u}$; the last inequality holds because $E_{m|u}$ commutes with its kernel projector and $\Id-E_{m|u}\ge0$. Summing over $m$ and using $\sum_mE_{m|u}=\Id$ gives $\sum_m\Pi_m\le(K-1)\Id$.

For each $m$, apply Lemma~\ref{lem:packing} to $T_{mt}$, $\Pi_mE_{m|t}\Pi_m$, and the common upper bound $\Pi_m\rho_A\Pi_m$, with $t\ne u$. This gives $\sum_{t\ne u}\tr(E_{m|t}T_{mt})\le\tr(\rho_A\Pi_m)$.

Summing over $m$, the left-hand side of Eq.~\eqref{eq:kbit} is at most $\tr[\rho_A\sum_m\Pi_m]\le K-1$. Each of its $N-1$ terms is at least the worst-case efficiency, which is therefore at most $(K-1)/(N-1)$.

In commuting-operator models, let $\psi$ be the shared unit vector, let $\{E_{m|s}\}_{m=0}^{K-1}$ be Alice's encoding POVMs, and let $D_{1|m,t}$ be Bob's acceptance effects. Define $v_{mt}$ and $V_{mt}$ as in the proof of Theorem~\ref{thm:sumrule}. Fix $u$, and let $\Pi_m$ project onto the closed span of $\{V_{mt}:t\ne u\}$. Since each of these subspaces lies in $\ker E_{m|u}$,
\[
\Pi_m\le\Id-E_{m|u},\qquad
\sum_m\Pi_m\le(K-1)\Id.
\]
For each fixed $m$, first test $\Pi_m$ and, on success, apply Alice's sequential tests with $t\ne u$. Their conclusive effects satisfy $\sum_{t\ne u}K_{mt}\le\Pi_m$. Equation~\eqref{eq:cqc-packing} therefore gives
\[
\sum_{t\ne u}\langle v_{mt},E_{m|t}v_{mt}\rangle
\le\langle\psi,\Pi_m\psi\rangle.
\]
Here the tests are constructed separately for each $m$; the projectors $\Pi_m$ need not be mutually orthogonal. Summing over $m$ yields
\[
\sum_{t\ne u}\Pr(G=1\mid t,t)
\le\sum_m\langle\psi,\Pi_m\psi\rangle
\le K-1,
\]
which completes the proof.\hfill$\square$

\paragraph{Communication cost.} We measure communication by the length of Alice's message and success by the worst case over input pairs. Message length is counted in the worst case, as the number of classical bits Alice sends; shared randomness and pre-shared correlations are free, and we do not consider quantum messages, expected message lengths or two-way communication. The quantum lower bounds below hold under the same conventions for arbitrary entanglement and in the commuting-operator model. For $\eta\in(0,1]$, let $C^\eta_{\Omega}(N)$ be the smallest $k$ for which some one-way zero-error protocol over $\Omega$ with a $k$-bit message satisfies $\min_{s,t}\Pr(G\neq\bm ?\,|\,s,t)\ge\eta$.

\begin{corollary}[Communication cost]\label{cor:cost}
For every $\eta\in(0,1]$,
\begin{equation}\label{eq:cost}
\big\lceil\log_2[1+\eta(N-1)]\big\rceil\le C^\eta_{\C_{\mathrm{qc}}}(N)\le C^\eta_{\C_{\mathrm L}}(N)\le\min\big\{\big\lceil\log_2(1+\eta N)\big\rceil,\,\lceil\log_2N\rceil\big\},
\end{equation}
whereas $C^{1/19}_\AQ(N)=1$. In particular, for $N=2^n$ inputs, attaining worst-case conclusive probability $1/19$ requires at least $n-\log_2 19$ bits with entanglement, but a single bit with almost-quantum correlations.
\end{corollary}

\begin{proof}
By Theorem~\ref{thm:kbit}, a $k$-bit protocol over $\C_{\mathrm{qc}}$ with worst-case conclusive probability $\eta$ needs $(2^k-1)/(N-1)\ge\eta$; since $k$ is an integer, this gives the left inequality in Eq.~\eqref{eq:cost}. The middle inequality holds because $\C_{\mathrm L}\subseteq\C_{\mathrm{qc}}$, and the right one follows from the classical protocol with $K=2^k$ messages (Alice sends $s$ itself if $K\ge N$, which gives the cap $\lceil\log_2N\rceil$). The outer bounds differ by at most one, because $\log_2(1+\eta N)-\log_2[1+\eta(N-1)]<1$. Theorem~\ref{thm:qstar} gives $C^{1/19}_\AQ(N)\le1$. Without communication, Bob's output distribution does not depend on $s$, so zero error forces $\Pr(G=1|t,t)=\Pr(G=1|s,t)=0$ for $s\neq t$; no protocol over $\C_{\mathrm{NS}}$ without communication has positive efficiency. Finally, $\log_2[1+(N-1)/19]\ge\log_2(N/19)=n-\log_2 19$.
\end{proof}

\noindent Hence, for every target $\eta$, entanglement, even in the commuting-operator model, shortens Alice's message by at most one bit relative to shared randomness. Almost-quantum correlations, by contrast, reduce the cost of reaching $\eta=1/19$ from at least $n-\log_2 19\approx n-4.25$ bits to a single bit.

\paragraph{Zero-error communication complexity.} In the standard zero-error (Las Vegas) model of communication complexity~\cite{MehlhornSchmidt1982,Klauck2000}, Bob may abstain with probability at most $1/2$ on every input pair, so the relevant one-way cost is $C^{1/2}_{\Omega}(N)$. The constant efficiency of almost-quantum correlations can then be amplified by repetition.

\begin{corollary}[Zero-error communication complexity]\label{cor:lasvegas}
For $N=2^n$, $C^{1/2}_{\C_{\mathrm L}}(N)=C^{1/2}_{\C_{\mathrm q}}(N)=C^{1/2}_{\C_{\mathrm{qc}}}(N)=n$, whereas $C^{1/2}_\AQ(N)\le13$. That is, one-way zero-error protocols for $\EQ_N$ that abstain with probability at most $1/2$ on every input pair require exactly $n$ bits with shared randomness, with entanglement and in the commuting-operator model, but at most $13$ bits with almost-quantum correlations.
\end{corollary}

\begin{proof}
For the lower bound, Eq.~\eqref{eq:cost} with $\eta=1/2$ gives $C^{1/2}_{\C_{\mathrm{qc}}}(N)\ge\lceil\log_2(2^{n-1}+1/2)\rceil=n$. For the upper bound, Alice can send $s$ itself, which costs $n$ bits, so $C^{1/2}_{\C_{\mathrm L}}(N)\le n$. Since $\C_{\mathrm L}\subseteq\C_{\mathrm q}\subseteq\C_{\mathrm{qc}}$, we have $n\le C^{1/2}_{\C_{\mathrm{qc}}}\le C^{1/2}_{\C_{\mathrm q}}\le C^{1/2}_{\C_{\mathrm L}}\le n$. With almost-quantum correlations, the parties run $13$ independent copies of the one-bit protocol of Theorem~\ref{thm:qstar}, each with its own copy of $\bm p^\star$ and its own shared randomness, so Alice sends $13$ bits. Each copy never errs and is conclusive with probability $1/19$ on every input pair, independently of the other copies. Bob outputs any conclusive answer. Hence he abstains with probability $(18/19)^{13}<0.496$; twelve copies would not suffice, since $(18/19)^{12}>0.52$. The joint resource is almost quantum because tensor products of AQ realizations satisfy Eq.~\eqref{eq:AQ} factor by factor (Lemma~\ref{lem:closure}(ii)).
\end{proof}

\noindent For comparison, the transcript-guessing bound of Ref.~\cite{KlauckdeWolf2013}, adapted to classical messages in Proposition~\ref{prop:guess}, gives $2^c\ge N/2$, i.e., at least $n-1$ bits, for entanglement-assisted and commuting-operator protocols that exchange $c$ classical bits, even interactively, and Klauck and de Wolf obtained a lower bound of $n/2-1/2$ when the messages are qubits~\cite{KlauckdeWolf2013}. Corollary~\ref{cor:lasvegas} sharpens the one-way bound for classical messages to exactly $n$ bits and extends it to commuting-operator models. Almost-quantum correlations need at most $13$ bits.

\section{Equality certification with vector correlations}\label{sec:vec}
This section proves part~(iii) of Theorem~\ref{thm:models} and the separation of $\C_\AQ$ from $\C_{\mathrm{Vec}}$ stated after it. For certificates of equality from binary-outcome boxes, $\mathtt{Level\,1}$ of the NPA hierarchy attains
\begin{equation}\label{eq:vec_main}
\eta^{\rm eq}_{\C_{\mathrm{Vec}}}(N)=\frac{N}{8(N-2)}\qquad(N\ge4),
\end{equation}
which exceeds the quantum value $1/N$ for every $N\ge5$ and decreases to $1/8$ (Proposition~\ref{prop:vec} below). The proof needs no semidefinite programming: the upper bound combines two inequalities, each expressing the nonnegativity of a squared norm, and an explicit realization attains it. Mixing the optimal level-1 certificate with the classical inequality certificate, as in Eq.~\eqref{eq:opt_twosided}, gives $\eta_{\C_{\mathrm{Vec}}}(N)\ge N/(10N-16)>1/10$ for $N\ge4$, which exceeds the quantum value $1/N$ for every $N\ge9$. For every $N\ge5$ the optimal level-1 box lies outside $\C_\AQ$, which yields an explicit, analytic separation $\C_\AQ\subsetneq\C_{\mathrm{Vec}}$. The difference stems from the vectors $A_xB_x\psi$ that the $\mathtt{1+AB}$ level requires and level~1 does not (Lemma~\ref{lem:gram}); they lower the attainable certificate from about $1/8$ to $1/17$.

For $N\ge5$, vector correlations also violate the witness~\eqref{eq:witness}. The optimal level-1 boxes have the same structure as $\bm p^\star$ and are listed in Table~\ref{tab:vec}.

\begin{proposition}[Equality certification with vector correlations]\label{prop:vec}
Let $N\ge3$, and let $\bm p\in\C_{\mathrm{Vec}}\cap\mathrm F_0$ be a binary-outcome box with $N$ settings per party. Then
\begin{equation}\label{eq:vec_sum}
\sum_{x=1}^Np(0,0|x,x)\le
\begin{cases}
1, & N\in\{3,4\},\\[2pt]
\dfrac{N^2}{8(N-2)}, & N\ge4,
\end{cases}
\end{equation}
where the two expressions agree at $N=4$, and each bound is attained. Consequently $\eta^{\rm eq}_{\C_{\mathrm{Vec}}}(3)=1/3$ and
\begin{equation}\label{eq:vec_value}
\eta^{\rm eq}_{\C_{\mathrm{Vec}}}(N)=\frac{N}{8(N-2)},\qquad
\eta^{\rm eq}_{\C_{\mathrm{Vec}}}(N)-\frac1N=\frac{(N-4)^2}{8N(N-2)}\qquad(N\ge4).
\end{equation}
In particular, $\eta^{\rm eq}_{\C_{\mathrm{Vec}}}(N)>1/N$ if and only if $N\ge5$, and $\eta^{\rm eq}_{\C_{\mathrm{Vec}}}(N)$ decreases monotonically to $1/8$. For $N\ge4$ the optimum is attained by the box $\bm p^{\mathrm{Vec}}_N$ of Table~\ref{tab:vec}.
\end{proposition}

\begin{proof}
Let $\psi$, $\{\xi_{a|x}\}$ and $\{\zeta_{b|y}\}$ be a vector realization of $\bm p$ as in Section~\ref{sec:models}, and write $a_x=\xi_{0|x}$, $b_x=\zeta_{0|x}$ and $q_x=p(0,0|x,x)$. The proof uses only three properties of these vectors:
\begin{enumerate}
\item[(P1)] $\ip{\psi}{a_x}=\|a_x\|^2$ and $\ip{\psi}{b_x}=\|b_x\|^2$, because $\xi_{0|x}\perp\xi_{1|x}$ and $\xi_{0|x}+\xi_{1|x}=\psi$ give $\ip{\psi}{a_x}=\ip{a_x+\xi_{1|x}}{a_x}=\|a_x\|^2$, and likewise for Bob;
\item[(P2)] $\ip{a_x}{b_y}=p(0,0|x,y)=q_x\,\delta_{xy}$, because $\bm p\in\mathrm F_0$;
\item[(P3)] $q_x\le\tfrac12\big(\|a_x\|^2+\|b_x\|^2\big)$, by the Cauchy--Schwarz inequality $q_x=\ip{a_x}{b_x}\le\|a_x\|\,\|b_x\|$ and $2uv\le u^2+v^2$.
\end{enumerate}
The inner products in (P1) and (P2) are real, so the argument below holds verbatim in a real or a complex Hilbert space. Put
\[
A=\sum_xa_x,\qquad B=\sum_xb_x,\qquad S=\sum_x\big(\|a_x\|^2+\|b_x\|^2\big),\qquad Q=\sum_xq_x ,
\]
so that the left-hand side of Eq.~\eqref{eq:vec_sum} is $Q$. By (P1) and (P2), $\ip{\psi}{A+B}=S$ and $\ip{A}{B}=Q$.

\noindent\emph{Upper bound.} Let $c_x=a_x-b_x$. By (P2), $\|c_x\|^2=\|a_x\|^2+\|b_x\|^2-2q_x$, hence $\sum_x\|c_x\|^2=S-2Q$; moreover $\sum_xc_x=A-B$ and $\|A-B\|^2=\|A\|^2+\|B\|^2-2Q$. Expanding the squares gives the identity
\[
\sum_{x<y}\|c_x-c_y\|^2=N\sum_x\|c_x\|^2-\Big\|\sum_xc_x\Big\|^2 ,
\]
whose left-hand side is nonnegative. Therefore
\begin{equation}\label{eq:vec_ineq1}
\|A\|^2+\|B\|^2\le NS-2(N-1)Q ,
\end{equation}
with equality if and only if all $c_x$ coincide. Next, for every real $\lambda$,
\[
0\le\big\|\psi-\lambda(A+B)\big\|^2=1-2\lambda S+\lambda^2\big(\|A\|^2+\|B\|^2+2Q\big),
\]
and inserting Eq.~\eqref{eq:vec_ineq1} into the last term gives
\begin{equation}\label{eq:vec_ineq2}
0\le1-\lambda(2-\lambda N)\,S-2\lambda^2(N-2)\,Q .
\end{equation}
For $N\ge4$, the choice $\lambda=2/N$ removes the term in $S$ and yields $Q\le N^2/[8(N-2)]$; equality requires equality in Eq.~\eqref{eq:vec_ineq1} together with $\psi=\tfrac2N(A+B)$. For $N\in\{3,4\}$, the choice $\lambda=1/2$ makes the coefficient of $S$ equal to $-(4-N)/4\le0$, so the lower bound $S\ge2Q$ from (P3) may be inserted, and Eq.~\eqref{eq:vec_ineq2} becomes
\[
0\le1-\frac{4-N}{2}\,Q-\frac{N-2}{2}\,Q=1-Q .
\]
This proves Eq.~\eqref{eq:vec_sum}. Since $\min_xq_x\le Q/N$, it also gives $\eta^{\rm eq}_{\C_{\mathrm{Vec}}}(3)\le1/3$ and $\eta^{\rm eq}_{\C_{\mathrm{Vec}}}(N)\le N/[8(N-2)]$ for $N\ge4$.

\noindent\emph{Attainability.} For $N=3$, the local box in the proof of Theorem~\ref{thm:optimal} lies in $\C_{\mathrm L}\cap\mathrm F_0\subseteq\C_{\mathrm{Vec}}\cap\mathrm F_0$ and has $p(0,0|x,x)=1/3$ for every $x$. For $N\ge4$, we construct a box that meets both equality conditions above, namely that all $c_x$ coincide and $\psi=\tfrac2N(A+B)$. Let $\psi$ and $e$ be orthonormal, let $g_1,\dots,g_N$ be orthonormal and orthogonal to $\psi$ and $e$, and put $f_x=g_x-\frac1N\sum_yg_y$, so that $\sum_xf_x=0$ and $\ip{f_x}{f_y}=\delta_{xy}-1/N$. For real $\sigma,\tau,\delta$ set
\[
a_x=\sigma\psi+\tau e+\delta f_x,\qquad b_x=\sigma\psi-\tau e+\delta f_x .
\]
Then $c_x=2\tau e$ for every $x$, and $A+B=2N\sigma\psi$, so $\psi=\tfrac2N(A+B)$ holds if and only if $\sigma=1/4$. The relevant inner products are
\begin{gather*}
\ip{\psi}{a_x}=\ip{\psi}{b_x}=\sigma,\qquad \|a_x\|^2=\|b_x\|^2=\sigma^2+\tau^2+\delta^2\,\frac{N-1}{N},\\
\ip{a_x}{b_y}=\sigma^2-\tau^2+\delta^2\Big(\delta_{xy}-\frac1N\Big).
\end{gather*}
Property (P2) for $x\neq y$ requires $\sigma^2-\tau^2=\delta^2/N$, and then $q:=\ip{a_x}{b_x}=\delta^2$. Property (P1) requires $\sigma=\sigma^2+\tau^2+\delta^2(N-1)/N$. With $\sigma=1/4$, these two conditions have the solution
\[
q=\delta^2=\frac{N}{8(N-2)},\qquad \tau^2=\frac1{16}-\frac{q}{N}=\frac{N-4}{16(N-2)},
\]
and $\tau$ is real if and only if $N\ge4$. Now take $\xi_{0|x}=a_x$, $\xi_{1|x}=\psi-a_x$, $\zeta_{0|y}=b_y$ and $\zeta_{1|y}=\psi-b_y$. By (P1), $\xi_{0|x}\perp\xi_{1|x}$ and $\zeta_{0|y}\perp\zeta_{1|y}$, and each pair sums to $\psi$. Since $\ip{\psi}{a_x}=\ip{\psi}{b_y}=\tfrac14$ and $\ip{a_x}{b_y}=q\,\delta_{xy}$, the inner products $p(a,b|x,y)=\ip{\xi_{a|x}}{\zeta_{b|y}}$ are
\begin{equation}\label{eq:vec_box}
\begin{gathered}
p(0,0|x,y)=q\,\delta_{xy},\qquad p(0,1|x,y)=\tfrac14-q\,\delta_{xy},\\
p(1,0|x,y)=\tfrac14-q\,\delta_{xy},\qquad p(1,1|x,y)=1-\tfrac14-\tfrac14+q\,\delta_{xy}=\tfrac12+q\,\delta_{xy}.
\end{gathered}
\end{equation}
With $q=N/[8(N-2)]$ this is the box $\bm p^{\mathrm{Vec}}_N$ of Table~\ref{tab:vec}. Its marginals $p_A(0|x)=p_B(0|y)=1/4$ do not depend on the settings, and all its entries are nonnegative because $\tfrac14-q=(N-4)/[8(N-2)]\ge0$. Hence $\bm p^{\mathrm{Vec}}_N\in\C_{\mathrm{Vec}}\cap\mathrm F_0$; it is realized in the $(N+1)$-dimensional space spanned by $\psi$, $e$ and the $f_x$. Since $p(0,0|x,x)=N/[8(N-2)]$ for every $x$, it attains the bound $Q\le N^2/[8(N-2)]$ and the worst-case bound $\eta^{\rm eq}_{\C_{\mathrm{Vec}}}(N)\le N/[8(N-2)]$. This proves that each bound in Eq.~\eqref{eq:vec_sum} is attained, and the first identity in Eq.~\eqref{eq:vec_value}. For $N=4$, Table~\ref{tab:vec} gives $p(0,0|x,x)=1/4$, $p(0,1|x,x)=p(1,0|x,x)=0$ and $p(1,1|x,x)=3/4$, so $\bm p^{\mathrm{Vec}}_4$ coincides with the local box in the proof of Theorem~\ref{thm:optimal}, and $Q=1$. As $N\to\infty$, the row $x=y$ of Table~\ref{tab:vec} tends to $(1/8,1/8,1/8,5/8)$, while the row $x\neq y$ does not depend on $N$.

\begin{table}[t]
\centering
\caption{\label{tab:vec}\textbf{The level-1 box $\bm p^{\mathrm{Vec}}_N(a,b|x,y)$.} Like $\bm p^\star$, it depends only on whether the settings coincide; its entries depend on $N$. It lies in $\C_{\mathrm{Vec}}\cap\mathrm F_0$ for every $N\ge4$, attains Proposition~\ref{prop:vec}, coincides with a local box for $N=4$ and lies outside $\C_\AQ$ for every $N\ge5$.}
\begin{tabular}{lcccc}
\toprule
 & $p(0,0)$ & $p(0,1)$ & $p(1,0)$ & $p(1,1)$\\
\midrule
$x=y$ & $N/[8(N-2)]$ & $(N-4)/[8(N-2)]$ & $(N-4)/[8(N-2)]$ & $(5N-8)/[8(N-2)]$\\
$x\neq y$ & $0$ & $1/4$ & $1/4$ & $1/2$\\
\bottomrule
\end{tabular}
\end{table}

\noindent\emph{Comparison with $1/N$.} For $N\ge4$,
\[
\frac{N}{8(N-2)}-\frac1N=\frac{N^2-8(N-2)}{8N(N-2)}=\frac{(N-4)^2}{8N(N-2)},
\]
which is positive if and only if $N\ge5$. Finally, $\eta^{\rm eq}_{\C_{\mathrm{Vec}}}(3)=1/3>1/4=\eta^{\rm eq}_{\C_{\mathrm{Vec}}}(4)$, $\frac{d}{dN}\frac{N}{8(N-2)}=-\frac{1}{4(N-2)^2}<0$ and $\frac{N}{8(N-2)}\to\frac18$, so $\eta^{\rm eq}_{\C_{\mathrm{Vec}}}(N)$ decreases monotonically to $1/8$.
\end{proof}

\noindent{\bf An explicit separation of $\C_\AQ$ from $\C_{\mathrm{Vec}}$.} For every $N\ge5$, the box $\bm p^{\mathrm{Vec}}_N$ of Table~\ref{tab:vec} is not almost quantum. If it were, it would be an AQ box on $\mathrm F_0$ with $\min_xp(0,0|x,x)=N/[8(N-2)]$, so $N/[8(N-2)]\le\eta^{\rm eq}_\AQ(N)$ by Eq.~\eqref{eq:opt_def}. This is false. For $5\le N\le8$, Theorem~\ref{thm:optimal} gives $\eta^{\rm eq}_\AQ(N)=1/N$, and $N/[8(N-2)]>1/N$ by Eq.~\eqref{eq:vec_value}. For $N\ge8$, $\eta^{\rm eq}_\AQ(N)=N^2/(17N^2-80N+64)$, and $N/[8(N-2)]>N^2/(17N^2-80N+64)$ is equivalent to $9N^2-64N+64>0$, which holds for every $N\ge6$. Hence $\{\bm p^{\mathrm{Vec}}_N\}_{N\ge5}$ is an explicit family in $\C_{\mathrm{Vec}}\setminus\C_\AQ$, which proves $\C_\AQ\subsetneq\C_{\mathrm{Vec}}$.
\par

\section{Discussion}\label{sec:discussion}

Almost-quantum correlations are known to satisfy nontrivial communication complexity, no advantage for nonlocal computation, macroscopic locality and local orthogonality, and numerical evidence suggested that they also satisfy information causality~\cite{Navascues2015}. Satisfying these principles does not, however, make them operationally close to quantum correlations. We have shown that a single bit of communication suffices to separate the two sets, with an efficiency ratio that grows without bound, in the zero-error certification of equality. For every Hilbert-space model of quantum correlations, from finite-dimensional tensor products to commuting operators, the optimal worst-case efficiency is $1/N$, whereas an explicit family of AQ boxes, one for each $N$, attains $1/19$. With longer messages the gap becomes a gap in communication cost: entanglement saves at most one bit over shared randomness, while AQ correlations attain a constant efficiency with a single bit and reduce the zero-error cost of deciding equality of $n$-bit strings from $n$ bits to at most $13$. For $N=2^n$ the efficiency ratio is exponential in $n$, while the communication costs are linear versus constant; neither statement implies that the two sets are far apart in a distance on correlations. Nor does the separation conflict with the fact that AQ correlations satisfy nontrivial communication complexity~\cite{Navascues2015}: equality is already cheap with bounded error and shared randomness, and the AQ advantage lies in certificates that are perfectly reliable yet arrive with a probability that does not shrink with $N$. The results of this paper do not by themselves bear on information causality; the Note added at the end of this section summarizes a companion work in which the boxes constructed here are shown to violate it.

In this task shared entanglement offers no advantage over shared randomness. Other tasks without quantum advantage are known, such as nonlocal computation~\cite{Linden2007} and the multipartite guess-your-neighbour's-input game~\cite{Almeida2010}. There, however, the classical bound follows from principles that AQ correlations also satisfy, namely no advantage for nonlocal computation and local orthogonality~\cite{Fritz2013,Navascues2015}, so AQ correlations offer no advantage either. In our task, quantum and classical correlations again coincide, but AQ correlations do strictly better, for every $N\ge14$ by Eq.~\eqref{eq:opt_twosided} with the optimal boxes; for certificates of equality from binary-outcome boxes, Theorem~\ref{thm:optimal} shows that they do so precisely when $N\ge9$. The obstruction can be traced to the packing lemma: its sequential argument composes several of Alice's operators and requires Bob's operations to commute with such products. Commutation as an operator identity, as in $\C_{\mathrm{qc}}$, guarantees this. The AQ condition~\eqref{eq:AQ}, by contrast, constrains only single operators acting on the state, and the box $\bm p^\star$ exploits exactly this freedom. This identifies a structural obstruction that separates $\C_{\mathrm{qc}}$ from $\C_\AQ$; satisfying the sum rule~\eqref{eq:sumrule} does not by itself characterize commuting-operator correlations, nor does it single out quantum theory. The same sum rule provides an exact-zero (face) witness of post-quantumness whose violation grows linearly in $N$.

It is instructive to compare this separation with earlier ones. That the quantum set is strictly contained in the AQ set was shown already in the simplest Bell scenario, with two settings and two outcomes per party, through a Bell expression whose minimum over quantum correlations lies slightly above $-1$, while an AQ correlation attains a lower value~\cite{Navascues2015}. Such Bell-functional separations, which can be recast as nonlocal games, establish that the two sets differ. Here the separation takes a different form: an explicit family, scalable in $N$, in a zero-error communication task, with a sharp quantum benchmark and an analytically determined binary-box optimum. On the face $\mathrm F_0$, the linear functional $\sum_xp(0,0|x,x)$ has the same maximum, $1$, over local, quantum and commuting-operator correlations. A violation of Eq.~\eqref{eq:witness} by a box lying exactly on $\mathrm F_0$ therefore excludes local and quantum explanations alike, and reveals an AQ advantage for an objective on which quantum correlations do not outperform shared randomness. Its maximum over AQ correlations equals $N\eta^{\rm eq}_\AQ(N)$, because the permutation symmetry used in Section~\ref{sec:optimal} lets us take all diagonal terms equal; by Theorem~\ref{thm:optimal} this is $N^3/(17N^2-80N+64)$ for $N\ge8$, which grows approximately as $N/17$. The ratio of the AQ maximum to the quantum maximum therefore grows without bound, much as the ratio of quantum to local maxima does for families of Bell inequalities with unbounded violation~\cite{JungePalazuelos2011,BRSW2012}, and here both maxima are known exactly. Over $\mathtt{Level\,1}$ of the NPA hierarchy the same functional reaches $N^2/[8(N-2)]$ for $N\ge4$ (Proposition~\ref{prop:vec}), so the constraints of the $\mathtt{1+AB}$ level lower the maximal violation by a factor approaching $17/8$. The analogy with Bell inequalities has a limitation: unlike a Bell inequality, Eq.~\eqref{eq:witness} holds only on the face $\mathrm F_0$, so its growing violation implies neither noise tolerance nor a quantitative distance from the quantum set (see the remark after Corollary~\ref{cor:witness}).

The separation lies between the adjacent sets $\C_{\mathrm{qc}}$ and $\C_\AQ$ and does not depend on the resolution of Tsirelson's problem. AQ correlations are also more than a convenient semidefinite relaxation. Dowker, Henson and Wallden showed~\cite{DowkerHensonWallden2014} that they coincide with the correlations generated by strongly positive decoherence functionals in Sorkin's quantum measure theory~\cite{Sorkin1994} when measurements may depend on earlier outcomes. This is a representation result under specified consistency assumptions, not a derivation from a particular model of quantum gravity. Quantum measure theory is a histories-based generalization of quantum mechanics developed with quantum gravity, and causal set theory in particular~\cite{Bombelli1987}, in mind. Any histories-based model that makes these AQ behaviours available as operational resources would therefore allow the protocols above, so the consistency conditions of such models do not by themselves enforce the quantum bound $1/N$ on zero-error certification of equality. Whether a specific theory of this kind realizes these behaviours physically is a separate question.

Several questions remain open. Our sharp quantum bounds concern one-way protocols. Interaction cannot remove the linear quantum lower bound: by the transcript-guessing argument of Ref.~\cite{KlauckdeWolf2013}, every entanglement-assisted or commuting-operator protocol that exchanges $c$ classical bits satisfies $\sum_t\Pr(G=1|t,t)\le2^c$ (Proposition~\ref{prop:guess}). What remains open is whether interaction can improve on the one-way values, a question about the constant rather than the linear scaling. For messages of $k\ge2$ bits, Theorem~\ref{thm:kbit} leaves a small gap between the quantum bound $(2^k-1)/(N-1)$ and the classical value $(2^k-1)/N$. The separation relies on zero error. This is intrinsic to the task, since with bounded error equality is already easy with shared randomness, but the witness~\eqref{eq:witness} is valid only on the face $\mathrm F_0$, and a quantitative version that tolerates nonzero off-diagonal coincidences, together with the precision it requires as a function of $N$, remains to be found. Whether a single AQ box can certify both equality and inequality more efficiently than the mixture~\eqref{eq:opt_twosided}, and whether boxes with more outcomes or message-dependent settings improve on Theorem~\ref{thm:optimal}, is also open. Finally, it would be interesting to determine whether the decay of zero-error certification efficiency with $N$ can be elevated to a physical principle that excludes AQ correlations at the level of Bell scenarios, in the spirit of nontrivial communication complexity~\cite{Brassard2006}, information causality~\cite{Pawlowski2009} and local orthogonality~\cite{Fritz2013}.

\paragraph{Note added.} In a companion work~\cite{DasBhowmikBanik2026}, two of present authors (ADB and MB) show that the AQ boxes $\bm p^\star$ of Theorem~\ref{thm:qstar} violate information causality~\cite{Pawlowski2009}, and do so without bound. The protocol uses a single copy of $\bm p^\star$ and one classical bit. Alice receives $kN$ independent fair bits, grouped into $N$ blocks of $k$ bits; she inputs into the box the index of the first block that consists only of ones, if there is one, and sends her outcome to Bob. Bob, asked about a bit of block $i$, inputs $y=i$ and guesses $1$ if the message and his outcome are both $0$, and $0$ otherwise. Because $p(0,0|x,y)=0$ for $x\neq y$, a guess of $1$ is never wrong; it therefore carries information linear in its probability, and it answers all $k$ questions about the block at once. The sum, over Alice's bits $Z_r$, of the mutual information $I(Z_r:G_r)$ between $Z_r$ and Bob's guess $G_r$, which information causality bounds by the number of communicated bits, is then at least $\tfrac{k}{17}\big[1-(1-2^{-k})^N\big]$. This exceeds $1$ for $k=18$ and $N=757\,693$, and it grows as $k(1-e^{-k})/17$ for $N=k2^k$. For a quantum box with $p(0,0|x,y)=s\,\delta_{xy}$, Corollary~\ref{cor:witness} gives $s\le1/N$, and the lower bound obtained from the same protocol is at most $k2^{-k}\le1/2$. With the level-1 boxes $\bm p^{\mathrm{Vec}}_N$ of Table~\ref{tab:vec}, the protocol violates information causality for every $N\ge1117$, in agreement with Ref.~\cite{Cavalcanti2010}. The violation thus rests on the property established here: zero-error certificates of equality whose rate, $1/17$ for $\bm p^\star$, does not decrease with $N$.

\appendix

\section{\texorpdfstring{An explicit local model for $\bm p^\star$}{An explicit local model for p*}}\label{app:local}
We now write $\bm p^\star$ explicitly as a mixture of deterministic strategies of the three types (i)--(iii) in the proof of Proposition~\ref{prop:local}. Type (iii) strategies supply the coincidences $p(0,0|x,x)$, while types (i) and (ii) raise Alice's and Bob's marginals to $3/17$ without creating any further coincidence. Let $\pi_N=4/(17-N)$, and mix the following three classes:
\begin{itemize}
\item \emph{Coincidences}, with probability $N/17$: a type-(iii) strategy with $x_0$ chosen uniformly from $[N]$. Alice outputs $0$ if and only if $x=x_0$, and Bob outputs $0$ if and only if $y=x_0$.
\item \emph{Extra zeros for Bob}, with probability $(17-N)/34$: a type-(i) strategy. Alice always outputs $1$, and Bob outputs $0$ if and only if $y\in F_B$, where each label is placed in $F_B$ independently with probability $\pi_N$.
\item \emph{Extra zeros for Alice}, with probability $(17-N)/34$: a type-(ii) strategy, the mirror image of the previous class. Bob always outputs $1$, and Alice outputs $0$ if and only if $x\in F_A$, with $F_A$ chosen in the same way.
\end{itemize}
The weights are nonnegative and sum to $N/17+(17-N)/17=1$. Here $\pi_N$ is not a distribution over labels but the bias of an independent coin tossed for each label; the induced distribution over subsets, $\Pr(F_B=S)=\pi_N^{|S|}(1-\pi_N)^{N-|S|}$, is normalized because $\sum_{S\subseteq[N]}\Pr(F_B=S)=(\pi_N+1-\pi_N)^N=1$. The only requirement is therefore $0\le\pi_N\le1$, which holds if and only if $N\le13$. Every strategy used lies on $\mathrm F_0$, so the mixture does too.

\noindent It remains to compute the box. Only the first class produces the outcome pair $(0,0)$, and it does so exactly when $x=y=x_0$, an event of probability $\delta_{xy}/N$. Hence
\[
p(0,0|x,y)=\frac{N}{17}\cdot\frac{\delta_{xy}}{N}=\frac{\delta_{xy}}{17}.
\]
Alice outputs $0$ in the first class when $x=x_0$ (probability $1/N$), never in the second, and in the third when $x\in F_A$ (probability $\pi_N$). Hence
\[
p_A(0|x)=\frac{N}{17}\cdot\frac1N+\frac{17-N}{34}\cdot0+\frac{17-N}{34}\cdot\frac{4}{17-N}=\frac1{17}+\frac2{17}=\frac3{17},
\]
and the same argument with the second and third classes interchanged gives $p_B(0|y)=3/17$. For binary outcomes these three numbers determine the whole box,
\begin{align*}
p(0,1|x,y)&=p_A(0|x)-p(0,0|x,y),\\
p(1,0|x,y)&=p_B(0|y)-p(0,0|x,y),\\
p(1,1|x,y)&=1-p_A(0|x)-p_B(0|y)+p(0,0|x,y),
\end{align*}
which reproduces Table~\ref{tab:qstar}. The mixture therefore equals $\bm p^\star$, so $\bm p^\star$ is local for every $N\le13$. The construction breaks down exactly where it must: for $N\ge14$, $\bm p^\star$ violates Eq.~\eqref{eq:localbound}, so no local model exists.

\section{Relation to the work of Klauck and de Wolf}\label{app:kdw}
The quantum part of our results is closely related to the work of Klauck and de Wolf~\cite{KlauckdeWolf2013}, who adapted the fooling-set method of classical communication complexity~\cite{Yao1979,KushilevitzNisan1997} to quantum protocols with prior entanglement. This appendix recalls their setting and results, relates their key inequality to Lemma~\ref{lem:packing} and to the witness of Corollary~\ref{cor:witness}, extends their argument to commuting-operator models and compares the resulting bounds with ours. Table~\ref{tab:kdw} summarizes the comparison.

\paragraph{Setting and results.} Let $f:X\times Y\to\{0,1\}$ be a Boolean function. A \emph{$1$-fooling set} of size $N$ is a set of $N$ input pairs which, after a relabelling of Bob's inputs, can be written as $\{(x,x)\}_{x\in[N]}$, such that $f(x,x)=1$ for every $x$ and, for $x\neq y$, $f(x,y)=0$ or $f(y,x)=0$. It is \emph{upper triangular} if, for some ordering of the labels, $f(x,y)=0$ whenever $x>y$. For equality all off-diagonal values vanish, so the diagonal is an upper-triangular $1$-fooling set of size $N$. A \emph{one-sided-error} protocol never outputs $1$ on an input with $f=0$, and outputs $1$ with probability at least $1/2$ on every input with $f=1$. For equality, such a protocol never certifies equality wrongly and certifies it with probability at least $1/2$ whenever the inputs coincide: it is the one-sided version of our task, in which equality is the only conclusive answer. Klauck and de Wolf denote by $Q_1^*(f)$ the one-sided-error communication complexity with unlimited prior entanglement, in which communication is counted in qubits and may be interactive. Their main result (Theorem~1 of Ref.~\cite{KlauckdeWolf2013}) states that
\begin{equation}\label{eq:kdw_main}
Q_1^*(f)\ \ge\ \tfrac12\log_2N-\tfrac12
\end{equation}
whenever $f$ has an upper-triangular $1$-fooling set of size $N$. For the $n$-bit equality, disjointness and greater-than functions, $N=2^n$, and Eq.~\eqref{eq:kdw_main} gives $n/2-1/2$, which superdense coding~\cite{BennettWiesner1992} shows to be optimal up to an additive constant. They also prove a bound with constant $1/4$ for arbitrary $1$-fooling sets, obtain lower bounds on zero-error (Las Vegas) complexity from $Q_0^*(f)\ge\max\{Q_1^*(f),Q_1^*(\neg f)\}$, and bound nondeterministic quantum communication complexity without shared resources in terms of fooling sets.

\paragraph{The two steps of the proof.} The proof of Eq.~\eqref{eq:kdw_main} has two steps. The first removes communication (Lemma~1 of Ref.~\cite{KlauckdeWolf2013}). Each communicated qubit is teleported~\cite{Bennett1993} with an additional EPR pair and two classical bits, and both parties proceed as if every pair of teleportation bits were $00$. These bits are uniformly random and independent of the inputs and of the rest of the protocol, so the resulting protocol without communication accepts, in the sense that both parties output $1$, with probability exactly $4^{-q}$ times the acceptance probability of the original $q$-qubit protocol. Applied to a one-sided-error protocol, it produces a tensor-product box that never accepts on the pairs $x>y$ and accepts each diagonal pair with probability at least $4^{-q}/2$. The second step bounds the diagonal by the following inequality (Claim~1 of Ref.~\cite{KlauckdeWolf2013}): if $w$ is a vector in a bipartite Hilbert space and $A_x$, $B_y$ are effects with $\ip{w}{(A_x\otimes B_y)w}=0$ for all $x>y$, then
\begin{equation}\label{eq:kdw_claim}
\sum_{x=1}^N\ip{w}{(A_x\otimes B_x)w}\le\|w\|^2 .
\end{equation}
Hence $N4^{-q}/2\le1$, which is Eq.~\eqref{eq:kdw_main}.

In the language of Section~\ref{sec:witness}, with acceptance relabelled as outcome $0$, Eq.~\eqref{eq:kdw_claim} is the face witness~\eqref{eq:witness} for tensor-product boxes, valid on the larger face
\[
\mathrm F_0^{\mathrm{ut}}=\big\{\bm p:\ p(0,0|x,y)=0\ \text{whenever}\ x>y\big\}\supseteq\mathrm F_0 .
\]
The two arguments run in opposite directions. Klauck and de Wolf convert a protocol into a box on this face and apply the inequality, whereas Corollary~\ref{cor:witness} converts a box on $\mathrm F_0$ into a one-bit protocol, protocol~(a) of Section~\ref{sec:qstar}, and applies the sum rule.

\paragraph{The inequality for commuting operators.} Klauck and de Wolf prove Eq.~\eqref{eq:kdw_claim} by induction on $N$, splitting $w$ with the support projection of $A_N\otimes\Id$. For tensor-product models it also follows from Lemma~\ref{lem:packing}, applied with the labels in decreasing order to $F_x=A_x$ and to Alice's steered operators $T_y=\tr_B\big[(\Id\otimes\sqrt{B_y})\,|w\rangle\langle w|\,(\Id\otimes\sqrt{B_y})\big]\le\tr_B|w\rangle\langle w|$: for $x>y$, $\tr(A_xT_y)=\ip{w}{(A_x\otimes B_y)w}=0$, and positivity gives $A_xT_y=0$. Both proofs rest on the same principle, that a test certain to fail on a state leaves the state undisturbed. The sequential form of Lemma~\ref{lem:packing} carries over to commuting-operator models, because the effects it constructs belong to the algebra generated by Alice's operators.

\begin{proposition}[Triangular packing]\label{prop:kdw}
Let $\psi$ be a vector in a Hilbert space, and let $A_1,\dots,A_N$ and $B_1,\dots,B_N$ be effects such that $A_xB_y=B_yA_x$ for all $x,y$, and $\ip{\psi}{A_xB_y\psi}=0$ whenever $x>y$. Then
\begin{equation}\label{eq:kdw_qc}
\sum_{x=1}^N\ip{\psi}{A_xB_x\psi}\le\|\psi\|^2 .
\end{equation}
In particular, every binary-outcome box $\bm p\in\C_{\mathrm{qc}}\cap\mathrm F_0^{\mathrm{ut}}$ satisfies $\sum_xp(0,0|x,x)\le1$, for every $N$.
\end{proposition}

\begin{proof}
Every operator in the norm-closed algebra generated by $A_1,\dots,A_N$, such as $\sqrt{\Id-A_x}$, commutes with every $B_y$ and with every $\sqrt{B_y}$. Put $v_y=\sqrt{B_y}\,\psi$. For $x>y$, $\|\sqrt{A_x}\,v_y\|^2=\ip{\psi}{A_xB_y\psi}=0$, so $A_xv_y=0$. Apply the L\"uders tests $A_N,A_{N-1},\dots,A_1$ in this order and stop at the first ``yes''. With $C_N=\Id$ and $C_{x-1}=\sqrt{\Id-A_x}\,C_x$, the outcome ``first yes at test $x$'' has effect $K_x=C_x^\dagger A_xC_x$, and, as in the proof of Lemma~\ref{lem:packing}, $\sum_xK_x=\Id-C_0^\dagger C_0\le\Id$. The tests performed before $A_y$ are those with $x>y$; they annihilate $v_y$, so $C_yv_y=v_y$ and $\ip{v_y}{A_yv_y}=\ip{v_y}{K_yv_y}$. Since $K_y$ commutes with $B_y$ and $0\le B_y\le\Id$, we have $K_y^{1/2}(\Id-B_y)K_y^{1/2}\ge0$, and therefore
\[
\ip{\psi}{A_yB_y\psi}=\ip{v_y}{A_yv_y}=\ip{v_y}{K_yv_y}=\ip{\psi}{K_yB_y\psi}\le\ip{\psi}{K_y\psi}.
\]
Summing over $y$ gives Eq.~\eqref{eq:kdw_qc}. For a box in $\C_{\mathrm{qc}}$, take a unit vector $\psi$ and commuting measurements that realize it, and put $A_x=A_{0|x}$ and $B_y=B_{0|y}$.
\end{proof}

\noindent Tensor-product effects $A_x\otimes\Id$ and $\Id\otimes B_y$ commute, so Proposition~\ref{prop:kdw} contains Eq.~\eqref{eq:kdw_claim}. On $\mathrm F_0$ it gives Corollary~\ref{cor:witness} for commuting-operator models without the restriction $N\ge3$, which enters Corollary~\ref{cor:witness} only through Theorem~\ref{thm:sumrule}. It fails for almost-quantum correlations: the boxes $\bm p^\star$ lie in $\C_\AQ\cap\mathrm F_0\subseteq\mathrm F_0^{\mathrm{ut}}$ and violate the bound for every $N\ge18$, and the optimal AQ boxes violate it for every $N\ge9$; at $\mathtt{Level\,1}$ the optimal boxes violate it for every $N\ge5$. The proof breaks where $K_y$, which composes several of Alice's operators, is moved past $B_y$; the AQ condition~\eqref{eq:AQ} provides commutation only for single operators acting on $\psi$. Read as a constraint on correlations, the inequality of Klauck and de Wolf thus holds throughout $\C_{\mathrm{qc}}$ and fails in $\C_\AQ$.

\paragraph{Removing classical communication.} When the parties communicate classical bits, the first step of the argument needs no teleportation. The following statement is the classical counterpart of Lemma~1 of Ref.~\cite{KlauckdeWolf2013}, combined with Proposition~\ref{prop:kdw}.

\begin{proposition}[Transcript guessing]\label{prop:guess}
Consider a protocol for $\EQ_N$, possibly interactive, in which the parties exchange $c$ classical bits in total, share randomness, and either share an entangled state or act on a common Hilbert space with commuting operator algebras. If the protocol never declares equality on an input pair with $s\neq t$, then
\begin{equation}\label{eq:guess}
\sum_{t=1}^N\Pr(G=1\,|\,t,t)\le2^c .
\end{equation}
\end{proposition}

\begin{proof}
Shared randomness can be absorbed into the shared state, and padding ensures that every run exchanges exactly $c$ bits. Let $r\in\{0,1\}^c$ be uniformly random, shared and independent of everything else. Without communicating, each party runs its part of the protocol on the shared state, using the bits of $r$ in place of the messages it would receive and checking whether the messages it would send agree with the corresponding bits of $r$. Alice outputs $0$ if all her checks succeed; Bob outputs $0$ if all his checks succeed and the run declares equality; otherwise a party outputs $1$. Each party's operations depend only on its input and on the messages it receives. Hence, for every transcript $\tau$, the probability that $r=\tau$, that all checks succeed and that the run declares equality equals $2^{-c}$ times the probability that a genuine run produces $\tau$ and declares equality. Summing over $\tau$ gives a box with $p(0,0|s,t)=2^{-c}\Pr(G=1|s,t)$. Each party's acceptance effect is built from its own operations alone, so the box is a tensor-product, respectively commuting-operator, box with binary outcomes, and by hypothesis it lies on $\mathrm F_0$. Proposition~\ref{prop:kdw} gives $2^{-c}\sum_t\Pr(G=1|t,t)\le1$.
\end{proof}

\paragraph{Comparison of the bounds.} For a single bit of one-way communication, Proposition~\ref{prop:guess} gives $\sum_t\Pr(G=1|t,t)\le2$ and hence $\eta_{\C_{\mathrm{qc}}}(N)\le2/N$. Theorem~\ref{thm:sumrule} removes the factor $2$, which is the price of guessing the message. It keeps the message instead: since $N\ge3$, a third label shows that the acceptance sectors of the two messages are orthogonal, so the packing for both messages fits into a single copy of Alice's reduced state, or of $\psi$, and the resulting value $1/N$ is attained with shared randomness. The restriction $N\ge3$ is essential: for $N=2$ Alice can send her label, the efficiency is $1$, and the bound $2$ of Eq.~\eqref{eq:guess} is attained. For $K$ one-way messages, the same argument with a uniformly random guess of the message gives $\sum_t\Pr(G=1|t,t)\le K$ and $\eta\le K/N$, whereas Theorem~\ref{thm:kbit} gives $\sum_{t\neq u}\Pr(G=1|t,t)\le K-1$ and $\eta\le(K-1)/(N-1)$, which is smaller precisely when $N>K$; the classical value is $(K-1)/N$. For the zero-error cost of $n$-bit equality, a protocol that is conclusive with probability at least $1/2$ on every input pair has $\sum_t\Pr(G=1|t,t)\ge N/2$, so Eq.~\eqref{eq:guess} requires $c\ge n-1$ classical bits, even with interaction, while Klauck and de Wolf obtain $n/2-1/2$ for qubit communication. Corollary~\ref{cor:lasvegas} determines the one-way cost with classical messages exactly, namely $n$ bits in every quantum model, whereas almost-quantum correlations need at most $13$. Quantum messages lie outside the scope of this paper: superdense coding halves their cost, and Eq.~\eqref{eq:kdw_main} is the relevant bound there.

\begin{table}[t]
\centering
\caption{\label{tab:kdw}\textbf{Comparison with Ref.~\cite{KlauckdeWolf2013}} for zero-error certification of equality of $N=2^n$ labels. ``Guessing'' refers to the removal of classical communication in Proposition~\ref{prop:guess}, the classical counterpart of their Lemma~1.}
\small\renewcommand{\arraystretch}{1.18}
\begin{tabular}{>{\raggedright\arraybackslash}p{3.2cm}>{\raggedright\arraybackslash}p{5.5cm}>{\raggedright\arraybackslash}p{5.5cm}}
\toprule
 & Klauck and de Wolf~\cite{KlauckdeWolf2013} & This work\\
\midrule
Functions & every $f$ with a large $1$-fooling set; sharp for upper-triangular ones & equality\\
Communication & interactive; qubits, or classical bits by guessing & one-way; $K$ classical messages\\
Correlations & finite-dimensional entanglement & $\C_{\mathrm L}$ to $\C_{\mathrm{qc}}$; also $\C_\AQ$ and $\C_{\mathrm{Vec}}$\\
Key inequality & Eq.~\eqref{eq:kdw_claim}: tensor products, zeros for $x>y$ & Lemma~\ref{lem:packing} and Proposition~\ref{prop:kdw}: commuting operators, zeros for $x>y$\\
One bit: $\sum_t\Pr(G=1|t,t)$ & $\le2$ by guessing, so $\eta\le2/N$ & $\le1$, so $\eta=1/N$ exactly (Theorem~\ref{thm:sumrule}, Corollary~\ref{cor:quantum})\\
$K$ messages: worst-case $\eta$ & $\le K/N$ by guessing & $\le(K-1)/(N-1)$ (Theorem~\ref{thm:kbit}); classical value $(K-1)/N$\\
Zero-error cost of $\EQ_N$ & $\ge n/2-1/2$ qubits; $\ge n-1$ classical bits by guessing; both with interaction & exactly $n$ classical bits one way (Corollary~\ref{cor:lasvegas}); at most $13$ with $\C_\AQ$\\
Post-quantum correlations & not considered & boxes in $\C_\AQ$ and $\C_{\mathrm{Vec}}$ violate the bound of Proposition~\ref{prop:kdw}\\
\bottomrule
\end{tabular}
\end{table}

\paragraph{The shared-randomness protocol.} Klauck and de Wolf also observe that shared randomness trivializes nondeterministic communication complexity: with a shared, uniformly random $r\in X$, Alice announces whether $x=r$, and Bob outputs $f(r,y)$ if she does and $0$ otherwise. For equality, this one-bit protocol never accepts an unequal pair and accepts every equal pair with probability $1/N$; it is the shared-randomness protocol of Corollary~\ref{cor:quantum}. Theorem~\ref{thm:sumrule} shows that no one-bit protocol assisted by entanglement, or by commuting-operator correlations, does better, whereas almost-quantum correlations do better by an unbounded factor (Theorems~\ref{thm:qstar} and~\ref{thm:optimal}).

\paragraph{Scope.} The results of Ref.~\cite{KlauckdeWolf2013} apply to every function with a large $1$-fooling set, to interactive protocols and to qubit communication, and they also bound nondeterministic quantum communication complexity. Our quantum bounds are restricted to equality and, apart from Proposition~\ref{prop:guess}, to one-way classical messages; within this setting they are exact, and they hold in the commuting-operator model. The post-quantum part of this paper has no counterpart in Ref.~\cite{KlauckdeWolf2013}. The contributions of this paper on the quantum side are therefore the exact one-way values (Theorems~\ref{thm:sumrule} and~\ref{thm:kbit} and Corollaries~\ref{cor:quantum}, \ref{cor:cost} and~\ref{cor:lasvegas}) and the extension of these bounds, together with Propositions~\ref{prop:kdw} and~\ref{prop:guess}, to commuting-operator models. The almost-quantum constructions and optimizations of Sections~\ref{sec:aq} and~\ref{sec:vec} do not rely on any of these bounds.
\par

\medskip
\noindent{\bf ACKNOWLEDGEMENT:} JK and BC acknowledge support from University Grants Commission, India (Reference no. 231620167137 and 241620129062, respectively). RKP acknowledges support from the EU (CHISTERA MoDIC) and from the National Research, Development and Innovation Office NKFIH (No. 2023-1.2.1-ERA-NET-2023-00009). MB acknowledges the financial support through the National Quantum Mission (NQM) of the Department of Science and Technology, Government of India.

\medskip
\noindent{\bf Use of AI tools.} The problem was conceived and all principal results were derived by the authors. A large language model (Claude, Anthropic) was used to assist with editing the text, simplifying the analytic certificates in Sections~\ref{sec:aq} and~\ref{sec:vec}. ChatGPT (OpenAI) was also used to assist with simplifying the quantum proofs in Section~\ref{sec:quantum}. Every derivation and certificate reported here was checked independently in exact arithmetic; no result relies on an assertion of correctness by an AI tool.\par


\bibliographystyle{unsrtnat}

\begin{thebibliography}{67}
\providecommand{\natexlab}[1]{#1}
\providecommand{\url}[1]{\texttt{#1}}
\expandafter\ifx\csname urlstyle\endcsname\relax
  \providecommand{\doi}[1]{doi: #1}\else
  \providecommand{\doi}{doi: \begingroup \urlstyle{rm}\Url}\fi

\bibitem[Cleve and Buhrman(1997)]{CleveBuhrman1997}
Richard Cleve and Harry Buhrman.
\newblock {Substituting quantum entanglement for communication}.
\newblock \emph{Phys. Rev. A}, 56:\penalty0 1201, 1997.
\newblock \doi{10.1103/PhysRevA.56.1201}.

\bibitem[Buhrman et~al.(2010)Buhrman, Cleve, Massar, and de~Wolf]{BuhrmanRMP2010}
Harry Buhrman, Richard Cleve, Serge Massar, and Ronald de~Wolf.
\newblock {Nonlocality and communication complexity}.
\newblock \emph{Rev. Mod. Phys.}, 82:\penalty0 665, 2010.
\newblock \doi{10.1103/RevModPhys.82.665}.

\bibitem[Buhrman et~al.(1998)Buhrman, Cleve, and Wigderson]{BuhrmanCleveWigderson1998}
Harry Buhrman, Richard Cleve, and Avi Wigderson.
\newblock {Quantum vs.\ classical communication and computation}.
\newblock In \emph{Proceedings of the 30th Annual ACM Symposium on Theory of Computing (STOC '98)}, pages 63--68, New York, 1998. ACM.
\newblock \doi{10.1145/276698.276713}.

\bibitem[Buhrman et~al.(2012)Buhrman, Regev, Scarpa, and de~Wolf]{BRSW2012}
Harry Buhrman, Oded Regev, Giannicola Scarpa, and Ronald de~Wolf.
\newblock {Near-optimal and explicit Bell inequality violations}.
\newblock \emph{Theory Comput.}, 8:\penalty0 623--645, 2012.
\newblock \doi{10.4086/toc.2012.v008a027}.

\bibitem[Chakraborty et~al.(2026)Chakraborty, Banik, and de~Wolf]{Chakraborty2026}
Ananya Chakraborty, Manik Banik, and Ronald de~Wolf.
\newblock {Exponential Advantage of Multipartite Entanglement over Quantum Communication with Applications to Bounded-Storage Cryptography}, 2026.
\newblock URL \url{https://arxiv.org/abs/2607.27957}.

\bibitem[Bell(1964)]{Bell1964}
John~S. Bell.
\newblock {On the Einstein Podolsky Rosen paradox}.
\newblock \emph{Physics Physique Fizika}, 1:\penalty0 195, 1964.
\newblock \doi{10.1103/PhysicsPhysiqueFizika.1.195}.

\bibitem[Bell(1966)]{Bell1966}
John~S. Bell.
\newblock {On the Problem of Hidden Variables in Quantum Mechanics}.
\newblock \emph{Rev. Mod. Phys.}, 38:\penalty0 447, 1966.
\newblock \doi{10.1103/RevModPhys.38.447}.

\bibitem[Mermin(1993)]{Mermin1993}
N.~David Mermin.
\newblock {Hidden variables and the two theorems of John Bell}.
\newblock \emph{Rev. Mod. Phys.}, 65:\penalty0 803, 1993.
\newblock \doi{10.1103/RevModPhys.65.803}.

\bibitem[Brunner et~al.(2014)Brunner, Cavalcanti, Pironio, Scarani, and Wehner]{Brunner2014}
Nicolas Brunner, Daniel Cavalcanti, Stefano Pironio, Valerio Scarani, and Stephanie Wehner.
\newblock {Bell nonlocality}.
\newblock \emph{Rev. Mod. Phys.}, 86:\penalty0 419, 2014.
\newblock \doi{10.1103/RevModPhys.86.419}.

\bibitem[Brukner et~al.(2004)Brukner, {\.Z}ukowski, Pan, and Zeilinger]{Brukner2004}
{\v C}aslav Brukner, Marek {\.Z}ukowski, Jian-Wei Pan, and Anton Zeilinger.
\newblock {Bell's inequalities and quantum communication complexity}.
\newblock \emph{Phys. Rev. Lett.}, 92:\penalty0 127901, 2004.
\newblock \doi{10.1103/PhysRevLett.92.127901}.

\bibitem[Popescu and Rohrlich(1994)]{PR1994}
Sandu Popescu and Daniel Rohrlich.
\newblock {Quantum nonlocality as an axiom}.
\newblock \emph{Found. Phys.}, 24:\penalty0 379, 1994.
\newblock \doi{10.1007/BF02058098}.

\bibitem[Popescu(2014)]{Popescu2014}
Sandu Popescu.
\newblock {Nonlocality beyond quantum mechanics}.
\newblock \emph{Nat. Phys.}, 10:\penalty0 264, 2014.
\newblock \doi{10.1038/nphys2916}.

\bibitem[van Dam(2013)]{vanDam2013}
Wim van Dam.
\newblock {Implausible consequences of superstrong nonlocality}.
\newblock \emph{Nat. Comput.}, 12:\penalty0 9, 2013.
\newblock \doi{10.1007/s11047-012-9353-6}.

\bibitem[Brassard et~al.(2006)Brassard, Buhrman, Linden, M{\'e}thot, Tapp, and Unger]{Brassard2006}
Gilles Brassard, Harry Buhrman, Noah Linden, Andr{\'e}~Allan M{\'e}thot, Alain Tapp, and Falk Unger.
\newblock {Limit on Nonlocality in Any World in Which Communication Complexity Is Not Trivial}.
\newblock \emph{Phys. Rev. Lett.}, 96:\penalty0 250401, 2006.
\newblock \doi{10.1103/PhysRevLett.96.250401}.

\bibitem[Brunner and Skrzypczyk(2009)]{BrunnerSkrzypczyk2009}
Nicolas Brunner and Paul Skrzypczyk.
\newblock {Nonlocality Distillation and Postquantum Theories with Trivial Communication Complexity}.
\newblock \emph{Phys. Rev. Lett.}, 102:\penalty0 160403, 2009.
\newblock \doi{10.1103/PhysRevLett.102.160403}.

\bibitem[Linden et~al.(2007)Linden, Popescu, Short, and Winter]{Linden2007}
Noah Linden, Sandu Popescu, Anthony~J. Short, and Andreas Winter.
\newblock {Quantum nonlocality and beyond: Limits from nonlocal computation}.
\newblock \emph{Phys. Rev. Lett.}, 99:\penalty0 180502, 2007.
\newblock \doi{10.1103/PhysRevLett.99.180502}.

\bibitem[Paw{\l}owski et~al.(2009)Paw{\l}owski, Paterek, Kaszlikowski, Scarani, Winter, and {\.Z}ukowski]{Pawlowski2009}
Marcin Paw{\l}owski, Tomasz Paterek, Dagomir Kaszlikowski, Valerio Scarani, Andreas Winter, and Marek {\.Z}ukowski.
\newblock {Information causality as a physical principle}.
\newblock \emph{Nature}, 461:\penalty0 1101, 2009.
\newblock \doi{10.1038/nature08400}.

\bibitem[Cavalcanti et~al.(2010)Cavalcanti, Salles, and Scarani]{Cavalcanti2010}
Daniel Cavalcanti, Alejo Salles, and Valerio Scarani.
\newblock {Macroscopically local correlations can violate information causality}.
\newblock \emph{Nat. Commun.}, 1:\penalty0 136, 2010.
\newblock \doi{10.1038/ncomms1138}.

\bibitem[Patra et~al.(2023)Patra, Naik, Lobo, Sen, Sidhardh, Alimuddin, and Banik]{Patra2023}
Ram~Krishna Patra, Sahil~Gopalkrishna Naik, Edwin~Peter Lobo, Samrat Sen, Govind~Lal Sidhardh, Mir Alimuddin, and Manik Banik.
\newblock {Principle of information causality rationalizes quantum composition}.
\newblock \emph{Phys. Rev. Lett.}, 130:\penalty0 110202, 2023.
\newblock \doi{10.1103/PhysRevLett.130.110202}.

\bibitem[Navascu{\'e}s and Wunderlich(2010)]{NavascuesWunderlich2010}
Miguel Navascu{\'e}s and Harald Wunderlich.
\newblock {A glance beyond the quantum model}.
\newblock \emph{Proc. R. Soc. A}, 466:\penalty0 881, 2010.
\newblock \doi{10.1098/rspa.2009.0453}.

\bibitem[Fritz et~al.(2013)Fritz, Sainz, Augusiak, Brask, Chaves, Leverrier, and Ac{\'i}n]{Fritz2013}
Tobias Fritz, Ana~Bel{\'e}n Sainz, Remigiusz Augusiak, Jonatan~Bohr Brask, Rafael Chaves, Anthony Leverrier, and Antonio Ac{\'i}n.
\newblock {Local orthogonality as a multipartite principle for quantum correlations}.
\newblock \emph{Nat. Commun.}, 4:\penalty0 2263, 2013.
\newblock \doi{10.1038/ncomms3263}.

\bibitem[Navascu{\'e}s et~al.(2015)Navascu{\'e}s, Guryanova, Hoban, and Ac{\'i}n]{Navascues2015}
Miguel Navascu{\'e}s, Yelena Guryanova, Matty~J. Hoban, and Antonio Ac{\'i}n.
\newblock {Almost quantum correlations}.
\newblock \emph{Nat. Commun.}, 6:\penalty0 6288, 2015.
\newblock \doi{10.1038/ncomms7288}.

\bibitem[Navascu{\'e}s et~al.(2007)Navascu{\'e}s, Pironio, and Ac{\'i}n]{NPA2007}
Miguel Navascu{\'e}s, Stefano Pironio, and Antonio Ac{\'i}n.
\newblock {Bounding the set of quantum correlations}.
\newblock \emph{Phys. Rev. Lett.}, 98:\penalty0 010401, 2007.
\newblock \doi{10.1103/PhysRevLett.98.010401}.

\bibitem[Navascu{\'e}s et~al.(2008)Navascu{\'e}s, Pironio, and Ac{\'i}n]{NPA2008}
Miguel Navascu{\'e}s, Stefano Pironio, and Antonio Ac{\'i}n.
\newblock {A convergent hierarchy of semidefinite programs characterizing the set of quantum correlations}.
\newblock \emph{New J. Phys.}, 10:\penalty0 073013, 2008.
\newblock \doi{10.1088/1367-2630/10/7/073013}.

\bibitem[Sorkin(1994)]{Sorkin1994}
Rafael~D. Sorkin.
\newblock {Quantum mechanics as quantum measure theory}.
\newblock \emph{Mod. Phys. Lett. A}, 9:\penalty0 3119, 1994.
\newblock \doi{10.1142/S021773239400294X}.

\bibitem[Dowker et~al.(2014)Dowker, Henson, and Wallden]{DowkerHensonWallden2014}
Fay Dowker, Joe Henson, and Petros Wallden.
\newblock {A histories perspective on characterizing quantum non-locality}.
\newblock \emph{New J. Phys.}, 16:\penalty0 033033, 2014.
\newblock \doi{10.1088/1367-2630/16/3/033033}.

\bibitem[Henson and Sainz(2015)]{HensonSainz2015}
Joe Henson and Ana~Bel{\'e}n Sainz.
\newblock {Macroscopic noncontextuality as a principle for almost-quantum correlations}.
\newblock \emph{Phys. Rev. A}, 91:\penalty0 042114, 2015.
\newblock \doi{10.1103/PhysRevA.91.042114}.

\bibitem[Gallego et~al.(2011)Gallego, W{\"u}rflinger, Ac{\'\i}n, and Navascu{\'e}s]{Gallego2011}
Rodrigo Gallego, Lars~Erik W{\"u}rflinger, Antonio Ac{\'\i}n, and Miguel Navascu{\'e}s.
\newblock {Quantum correlations require multipartite information principles}.
\newblock \emph{Phys. Rev. Lett.}, 107:\penalty0 210403, 2011.
\newblock \doi{10.1103/PhysRevLett.107.210403}.

\bibitem[Gonda et~al.(2018)Gonda, Kunjwal, Schmid, Wolfe, and Sainz]{Gonda2018}
Tom{\'a}{\v s} Gonda, Ravi Kunjwal, David Schmid, Elie Wolfe, and Ana~Bel{\'e}n Sainz.
\newblock {Almost quantum correlations are inconsistent with Specker's principle}.
\newblock \emph{Quantum}, 2:\penalty0 87, 2018.
\newblock \doi{10.22331/q-2018-08-27-87}.

\bibitem[Sainz et~al.(2018)Sainz, Guryanova, Ac{\'\i}n, and Navascu{\'e}s]{Sainz2018}
Ana~Bel{\'e}n Sainz, Yelena Guryanova, Antonio Ac{\'\i}n, and Miguel Navascu{\'e}s.
\newblock {Almost-quantum correlations violate the no-restriction hypothesis}.
\newblock \emph{Phys. Rev. Lett.}, 120:\penalty0 200402, 2018.
\newblock \doi{10.1103/PhysRevLett.120.200402}.

\bibitem[Das~Bhowmik and Banik(2026)]{DasBhowmikBanik2026}
Anandamay Das~Bhowmik and Manik Banik.
\newblock {Almost-Quantum Correlations Violate Information Causality}, 2026.
\newblock [companion paper].

\bibitem[Klauck and de~Wolf(2013)]{KlauckdeWolf2013}
Hartmut Klauck and Ronald de~Wolf.
\newblock {Fooling one-sided quantum protocols}.
\newblock In \emph{30th International Symposium on Theoretical Aspects of Computer Science (STACS 2013)}, volume~20 of \emph{Leibniz International Proceedings in Informatics (LIPIcs)}, pages 424--433, Dagstuhl, 2013. Schloss Dagstuhl--Leibniz-Zentrum f{\"u}r Informatik.
\newblock \doi{10.4230/LIPIcs.STACS.2013.424}.
\newblock \href{https://arxiv.org/abs/1204.4619}{arXiv:1204.4619}.

\bibitem[Tsirelson(1993)]{Tsirelson1993}
Boris~S. Tsirelson.
\newblock {Some results and problems on quantum Bell-type inequalities}.
\newblock \emph{Hadronic J. Suppl.}, 8:\penalty0 329, 1993.

\bibitem[Fritz(2012)]{Fritz2012}
Tobias Fritz.
\newblock {Tsirelson's problem and Kirchberg's conjecture}.
\newblock \emph{Rev. Math. Phys.}, 24:\penalty0 1250012, 2012.
\newblock \doi{10.1142/S0129055X12500122}.

\bibitem[Junge et~al.(2011)Junge, Navascu{\'e}s, Palazuelos, P{\'e}rez-Garc{\'i}a, Scholz, and Werner]{Junge2011}
Marius Junge, Miguel Navascu{\'e}s, Carlos Palazuelos, David P{\'e}rez-Garc{\'i}a, Volkher~B. Scholz, and Reinhard~F. Werner.
\newblock {Connes' embedding problem and Tsirelson's problem}.
\newblock \emph{J. Math. Phys.}, 52:\penalty0 012102, 2011.
\newblock \doi{10.1063/1.3514538}.

\bibitem[Dykema et~al.(2017)Dykema, Paulsen, and Prakash]{DykemaPaulsenPrakash2017}
Ken Dykema, Vern~I. Paulsen, and Jitendra Prakash.
\newblock {The Delta game}.
\newblock \href{https://arxiv.org/abs/1707.06186}{arXiv:1707.06186} [math.OA], 2017.

\bibitem[Reichenbach(1956)]{Reichenbach1956}
Hans Reichenbach.
\newblock \emph{{The Direction of Time}}.
\newblock University of California Press, Berkeley, 1956.

\bibitem[Bell(1990)]{Bell1990}
John~S. Bell.
\newblock {La nouvelle cuisine}.
\newblock In A.~Sarlemijn and P.~Kroes, editors, \emph{Between Science and Technology}, page~97. Elsevier, Amsterdam, 1990.
\newblock \doi{10.1016/B978-0-444-88659-0.50010-7}.

\bibitem[Wood and Spekkens(2015)]{WoodSpekkens2015}
Christopher~J. Wood and Robert~W. Spekkens.
\newblock {The lesson of causal discovery algorithms for quantum correlations: causal explanations of Bell-inequality violations require fine-tuning}.
\newblock \emph{New J. Phys.}, 17:\penalty0 033002, 2015.
\newblock \doi{10.1088/1367-2630/17/3/033002}.

\bibitem[Haag and Kastler(1964)]{HaagKastler1964}
Rudolf Haag and Daniel Kastler.
\newblock {An algebraic approach to quantum field theory}.
\newblock \emph{J. Math. Phys.}, 5:\penalty0 848, 1964.
\newblock \doi{10.1063/1.1704187}.

\bibitem[Coladangelo and Stark(2020)]{ColadangeloStark2020}
Andrea Coladangelo and Jalex Stark.
\newblock {An inherently infinite-dimensional quantum correlation}.
\newblock \emph{Nat. Commun.}, 11:\penalty0 3335, 2020.
\newblock \doi{10.1038/s41467-020-17077-9}.

\bibitem[Slofstra(2019)]{Slofstra2019}
William Slofstra.
\newblock {The set of quantum correlations is not closed}.
\newblock \emph{Forum Math. Pi}, 7:\penalty0 e1, 2019.
\newblock \doi{10.1017/fmp.2018.3}.

\bibitem[Scholz and Werner(2008)]{ScholzWerner2008}
Volkher~B. Scholz and Reinhard~F. Werner.
\newblock {Tsirelson's problem}.
\newblock \href{https://arxiv.org/abs/0812.4305}{arXiv:0812.4305} [math-ph], 2008.

\bibitem[Ozawa(2013)]{Ozawa2013}
Narutaka Ozawa.
\newblock {About the Connes embedding conjecture: algebraic approaches}.
\newblock \emph{Jpn. J. Math.}, 8:\penalty0 147, 2013.
\newblock \doi{10.1007/s11537-013-1280-5}.

\bibitem[Ji et~al.(2020)Ji, Natarajan, Vidick, Wright, and Yuen]{MIPRE2020}
Zhengfeng Ji, Anand Natarajan, Thomas Vidick, John Wright, and Henry Yuen.
\newblock {MIP*=RE}.
\newblock \href{https://arxiv.org/abs/2001.04383}{arXiv:2001.04383} [quant-ph], 2020.

\bibitem[Cirel'son(1980)]{Tsirelson1980}
Boris~S. Cirel'son.
\newblock {Quantum generalizations of Bell's inequality}.
\newblock \emph{Lett. Math. Phys.}, 4:\penalty0 93, 1980.
\newblock \doi{10.1007/BF00417500}.

\bibitem[Ivanovic(1987)]{Ivanovic1987}
I.~D. Ivanovic.
\newblock {How to differentiate between non-orthogonal states}.
\newblock \emph{Phys. Lett. A}, 123:\penalty0 257, 1987.
\newblock \doi{10.1016/0375-9601(87)90222-2}.

\bibitem[Dieks(1988)]{Dieks1988}
D.~Dieks.
\newblock {Overlap and distinguishability of quantum states}.
\newblock \emph{Phys. Lett. A}, 126:\penalty0 303, 1988.
\newblock \doi{10.1016/0375-9601(88)90840-7}.

\bibitem[Peres(1988)]{Peres1988}
Asher Peres.
\newblock {How to differentiate between non-orthogonal states}.
\newblock \emph{Phys. Lett. A}, 128:\penalty0 19, 1988.
\newblock \doi{10.1016/0375-9601(88)91034-1}.

\bibitem[Yao(1979)]{Yao1979}
Andrew Chi-Chih Yao.
\newblock {Some complexity questions related to distributive computing}.
\newblock In \emph{Proceedings of the 11th Annual ACM Symposium on Theory of Computing (STOC '79)}, pages 209--213, New York, 1979. ACM.
\newblock \doi{10.1145/800135.804414}.

\bibitem[Kushilevitz and Nisan(1997)]{KushilevitzNisan1997}
Eyal Kushilevitz and Noam Nisan.
\newblock \emph{{Communication Complexity}}.
\newblock Cambridge University Press, Cambridge, 1997.
\newblock \doi{10.1017/CBO9780511574948}.

\bibitem[Newman(1991)]{Newman1991}
Ilan Newman.
\newblock {Private vs. common random bits in communication complexity}.
\newblock \emph{Inf. Process. Lett.}, 39:\penalty0 67, 1991.
\newblock \doi{10.1016/0020-0190(91)90157-D}.

\bibitem[Buhrman et~al.(2001)Buhrman, Cleve, Watrous, and de~Wolf]{Buhrman2001}
Harry Buhrman, Richard Cleve, John Watrous, and Ronald de~Wolf.
\newblock {Quantum fingerprinting}.
\newblock \emph{Phys. Rev. Lett.}, 87:\penalty0 167902, 2001.
\newblock \doi{10.1103/PhysRevLett.87.167902}.

\bibitem[Ambainis(1996)]{Ambainis1996}
Andris Ambainis.
\newblock {Communication complexity in a 3-computer model}.
\newblock \emph{Algorithmica}, 16:\penalty0 298, 1996.
\newblock \doi{10.1007/BF01955678}.

\bibitem[Newman and Szegedy(1996)]{NewmanSzegedy1996}
Ilan Newman and Mario Szegedy.
\newblock {Public vs. private coin flips in one round communication games}.
\newblock In \emph{Proceedings of the 28th Annual ACM Symposium on Theory of Computing (STOC '96)}, pages 561--570, New York, 1996. ACM.
\newblock \doi{10.1145/237814.238004}.

\bibitem[Mehlhorn and Schmidt(1982)]{MehlhornSchmidt1982}
Kurt Mehlhorn and Erik~M. Schmidt.
\newblock {Las Vegas is better than determinism in VLSI and distributed computing}.
\newblock In \emph{Proceedings of the 14th Annual ACM Symposium on Theory of Computing (STOC '82)}, pages 330--337, New York, 1982. ACM.
\newblock \doi{10.1145/800070.802208}.

\bibitem[Klauck(2000)]{Klauck2000}
Hartmut Klauck.
\newblock {On quantum and probabilistic communication: Las Vegas and one-way protocols}.
\newblock In \emph{Proceedings of the 32nd Annual ACM Symposium on Theory of Computing (STOC '00)}, pages 644--651, New York, 2000. ACM.
\newblock \doi{10.1145/335305.335396}.

\bibitem[Winter(1999)]{Winter1999}
Andreas Winter.
\newblock {Coding theorem and strong converse for quantum channels}.
\newblock \emph{IEEE Trans. Inf. Theory}, 45:\penalty0 2481, 1999.
\newblock \doi{10.1109/18.796385}.

\bibitem[Sen(2012)]{Sen2012}
Pranab Sen.
\newblock {Achieving the Han-Kobayashi inner bound for the quantum interference channel}.
\newblock In \emph{Proceedings of the IEEE International Symposium on Information Theory (ISIT)}, pages 736--740, 2012.
\newblock \doi{10.1109/ISIT.2012.6284656}.
\newblock \href{https://arxiv.org/abs/1109.0802}{arXiv:1109.0802}.

\bibitem[Wilde(2013)]{Wilde2013}
Mark~M. Wilde.
\newblock {Sequential decoding of a general classical-quantum channel}.
\newblock \emph{Proc. R. Soc. A}, 469:\penalty0 20130259, 2013.
\newblock \doi{10.1098/rspa.2013.0259}.

\bibitem[Gao(2015)]{Gao2015}
Jingliang Gao.
\newblock {Quantum union bounds for sequential projective measurements}.
\newblock \emph{Phys. Rev. A}, 92:\penalty0 052331, 2015.
\newblock \doi{10.1103/PhysRevA.92.052331}.

\bibitem[Heinosaari and Wolf(2010)]{HeinosaariWolf2010}
Teiko Heinosaari and Michael~M. Wolf.
\newblock {Nondisturbing quantum measurements}.
\newblock \emph{J. Math. Phys.}, 51:\penalty0 092201, 2010.
\newblock \doi{10.1063/1.3480658}.

\bibitem[Almeida et~al.(2010)Almeida, Bancal, Brunner, Ac{\'\i}n, Gisin, and Pironio]{Almeida2010}
Mafalda~L. Almeida, Jean-Daniel Bancal, Nicolas Brunner, Antonio Ac{\'\i}n, Nicolas Gisin, and Stefano Pironio.
\newblock {Guess your neighbor's input: A multipartite nonlocal game with no quantum advantage}.
\newblock \emph{Phys. Rev. Lett.}, 104:\penalty0 230404, 2010.
\newblock \doi{10.1103/PhysRevLett.104.230404}.

\bibitem[Junge and Palazuelos(2011)]{JungePalazuelos2011}
Marius Junge and Carlos Palazuelos.
\newblock {Large violation of Bell inequalities with low entanglement}.
\newblock \emph{Commun. Math. Phys.}, 306:\penalty0 695--746, 2011.
\newblock \doi{10.1007/s00220-011-1296-8}.

\bibitem[Bombelli et~al.(1987)Bombelli, Lee, Meyer, and Sorkin]{Bombelli1987}
Luca Bombelli, Joohan Lee, David Meyer, and Rafael~D. Sorkin.
\newblock {Space-time as a causal set}.
\newblock \emph{Phys. Rev. Lett.}, 59:\penalty0 521, 1987.
\newblock \doi{10.1103/PhysRevLett.59.521}.

\bibitem[Bennett and Wiesner(1992)]{BennettWiesner1992}
Charles~H. Bennett and Stephen~J. Wiesner.
\newblock {Communication via one- and two-particle operators on Einstein-Podolsky-Rosen states}.
\newblock \emph{Phys. Rev. Lett.}, 69:\penalty0 2881--2884, 1992.
\newblock \doi{10.1103/PhysRevLett.69.2881}.

\bibitem[Bennett et~al.(1993)Bennett, Brassard, Cr{\'e}peau, Jozsa, Peres, and Wootters]{Bennett1993}
Charles~H. Bennett, Gilles Brassard, Claude Cr{\'e}peau, Richard Jozsa, Asher Peres, and William~K. Wootters.
\newblock {Teleporting an unknown quantum state via dual classical and Einstein-Podolsky-Rosen channels}.
\newblock \emph{Phys. Rev. Lett.}, 70:\penalty0 1895--1899, 1993.
\newblock \doi{10.1103/PhysRevLett.70.1895}.

\end{thebibliography}

\end{document}